\documentclass{article}

\usepackage{booktabs}       
\usepackage{amsfonts}       
\usepackage[american]{babel}
\usepackage{csquotes}
\usepackage{longtable}
\usepackage{multirow,setspace}
\usepackage{pdfpages}
\usepackage{tikz}
\usepackage{array}
\usepackage{amsmath}
\usepackage{amssymb}
\usepackage{mathtools}
\usepackage[flushleft]{threeparttable}
\usepackage{bbm} 
\usepackage{siunitx} 
\usepackage{etoolbox} 
\usepackage{makecell} 
\usepackage{calc} 
\usepackage{float} 
\usetikzlibrary{positioning}
\usepackage{algorithm} 
\usepackage{authblk} 
\usepackage{mathptmx} 
\usepackage{indentfirst} 
\usepackage[margin=1.25in]{geometry} 
\PassOptionsToPackage{hyphens}{url}\usepackage[hidelinks]{hyperref}

\newcommand\blfootnote[1]{%
  \begingroup
  \renewcommand\thefootnote{}\footnote{#1}%
  \addtocounter{footnote}{-1}%
  \endgroup
}

\renewenvironment{abstract}
 {\small
  \begin{center}
  \bfseries \abstractname\vspace{-.5em}\vspace{0pt}
  \end{center}
  \list{}{%
    \setlength{\leftmargin}{.5in}
    \setlength{\rightmargin}{\leftmargin}%
  }%
  \item\relax}
 {\endlist}

\robustify\bfseries

\makeatletter
\let\pgfmathrandomX=\pgfmathrandom@
\usepackage{pgfplots}%
\let\pgfmathrandom@=\pgfmathrandomX
\makeatother
\pgfplotsset{compat=1.14}
\usepackage{ifthen}

\usetikzlibrary{shapes,arrows,positioning,calc,chains,scopes}

\definecolor{snowymint}{HTML}{E3F8D1}
\definecolor{wepeep}{HTML}{FAD2D2}
\definecolor{portafino}{HTML}{F5EE9D}
\definecolor{plum}{HTML}{DCACEF}
\definecolor{sail}{HTML}{A3CEEE}
\definecolor{highland}{HTML}{6D885A}

\tikzstyle{signal}=[arrows={-latex},draw=black,line width=1.5pt,rounded corners=4pt]

\tikzstyle{block}=[draw=black,line width=1.0pt]
\tikzstyle{cell}=[style=block,draw=highland,fill=snowymint,
    rounded corners]
\tikzstyle{celllayer}=[style=block,draw,fill=portafino,
    inner sep=1pt,outer sep=0,
    minimum width=28pt, minimum height=14pt]
\tikzstyle{pointwise}=[style=block,ellipse,fill=wepeep,
    inner sep=1pt,outer sep=0, minimum size=12pt]

\tikzstyle{netnode}=[circle, inner sep=0pt, text width=22pt, align=center, line width=1.0pt]
\tikzstyle{inputnode}=[netnode, fill=white,draw=black]
\tikzstyle{hiddennode}=[netnode, fill=white,draw=black]
\tikzstyle{outputnode}=[netnode, fill=white,draw=black]

\def\layerwidth{90pt}
\def\layerheight{14pt}

\tikzstyle{layer}=[style=block, draw, fill=black!20!white,
    inner sep=1pt,outer sep=0, font=\footnotesize,
    text centered, 
    minimum width=\layerwidth, minimum height=\layerheight]

\tikzstyle{fc}=[style=layer, fill=blue!30!white]
\tikzstyle{conv}=[style=layer, fill=green!30!white]
\tikzstyle{activation}=[style=layer, fill=orange!30!white]
\tikzstyle{pool}=[style=layer, fill=red!30!white]
\tikzstyle{bn}=[style=layer, fill=cyan!30!white]
\tikzstyle{recurrent}=[style=layer, fill=purple!30!white]
\tikzstyle{softmax}=[style=layer, fill=yellow!30!white]
\tikzstyle{point}=[]
\tikzstyle{branch}=[coordinate]

\def\vlayerwidth{30pt}
\def\vlayerheight{3pt}
\def\vblockheight{28pt}

\tikzstyle{vlayer}=[minimum width=\vlayerwidth, minimum height=\vlayerheight]
\tikzstyle{vblock}=[minimum width=\vlayerwidth, minimum height=\vblockheight, text width=1cm, align=center]

\colorlet{fn}{gray!90!green!30!white}
\colorlet{tp}{green!40!white}
\colorlet{fp}{red!40!white}
\colorlet{tn}{gray!90!red!20!white}

\newlength{\minuslen}
\newlength{\zerolen}

\newcommand{\algcomment}[1]{%
    \vspace{-\baselineskip}%
    \noindent%
    {\footnotesize #1\par}%
    \vspace{\baselineskip}%
    }

\DeclareMathOperator*{\argmin}{arg\,min}
\DeclareMathOperator*{\argmax}{arg\,max}
\DeclareMathOperator{\dkl}{\textit{D}_{KL}}
\DeclareMathOperator{\elbo}{ELBO}
\DeclareMathOperator{\iwelbo}{IW-ELBO}

\DeclareMathOperator{\fnn}{FNN}

\DeclareMathOperator{\bias}{bias}

\title{\textbf{A Deep Learning Algorithm for High-Dimensional Exploratory Item Factor Analysis\thanks{This is a preprint of an article published in Psychometrika. The final authenticated version is
available online at: \url{https://doi.org/doi:\%2010.1007/s11336-021-09748-3}}}}
\author{Christopher J. Urban\thanks{Correspondence to \texttt{cjurban@live.unc.edu}.} }
\author{Daniel J. Bauer}

\affil{L.L. Thurstone Psychometric Laboratory in the
Department of Psychology and Neuroscience \\
University of North Carolina at Chapel Hill}

\date{\vspace{-2em}}

\begin{document}

\maketitle

\begin{abstract}
\quad Marginal maximum likelihood\blfootnote{This material is based upon work supported by the National Science Foundation Graduate
Research Fellowship under Grant No. DGE-1650116.} (MML) estimation is the preferred approach to fitting item response theory models in psychometrics due to the MML estimator's consistency, normality, and efficiency as the sample size tends to infinity. However, state-of-the-art MML estimation procedures such as the Metropolis-Hastings Robbins-Monro (MH-RM) algorithm as well as approximate MML estimation procedures such as variational inference (VI) are computationally time-consuming when the sample size and the number of latent factors are very large. In this work, we investigate a deep learning-based VI algorithm for exploratory item factor analysis (IFA) that is computationally fast even in large data sets with many latent factors. The proposed approach applies a deep artificial neural network model called an importance-weighted autoencoder (IWAE) for exploratory IFA. The IWAE approximates the MML estimator using an importance sampling technique wherein increasing the number of importance-weighted (IW) samples drawn during fitting improves the approximation, typically at the cost of decreased computational efficiency. We provide a real data application that recovers results aligning with psychological theory across random starts. Via simulation studies, we show that the IWAE yields more accurate estimates as either the sample size or the number of IW samples increases (although factor correlation and intercepts estimates exhibit some bias) and obtains similar results to MH-RM in less time. Our simulations also suggest that the proposed approach performs similarly to and is potentially faster than constrained joint maximum likelihood estimation, a fast procedure that is consistent when the sample size and the number of items simultaneously tend to infinity.

\vspace{.5em}Key words: Deep learning, artificial neural network, variational inference, variational autoencoder, importance sampling, importance weighted autoencoder, item response theory, categorical factor analysis, latent variable modeling
\end{abstract}

\section{Introduction}

Psychology and education researchers often collect large-scale test data with many respondents and many items in order to measure unobserved latent constructs such as personality traits or cognitive abilities. When test items are dichotomous (e.g., ``Yes'' or ``No'') or polytomous (e.g., ``Always'', ``Frequently'', ``Occasionally'', or ``Never''), item factor analysis (IFA) is a principled alternative to linear factor analysis for summarizing the items using a smaller number of continuous latent factors. Exploratory IFA (Bock et al., 1988) in particular is an indispensable tool for uncovering the latent structure underlying a test by estimating the associations between items and latent factors (i.e., the factor loadings) in a data-driven manner. See Bolt (2005) or Wirth and Edwards (2007) for overviews of exploratory IFA.

Exploratory IFA parameters are most often estimated using Bock and Aitkin's (1981) marginal maximum likelihood (MML) estimator, which enjoys consistency, normality, and efficiency as the sample size approaches infinity. The MML approach estimates the item parameters by maximizing the marginal likelihood of the observed item responses, which is obtained by integrating out the latent factors. Problematically, the computational complexity of evaluating this integral is exponential in the dimension of the latent space, making direct evaluation of the marginal likelihood computationally burdensome in the high-dimensional setting. A variety of methods for approximating the integrals have been proposed, including adaptive Gaussian quadrature (Rabe-Hesketh et al., 2005; Schilling \& Bock, 2005), Laplace approximation (e.g., Huber et al., 2004), Monte Carlo integration (e.g., Meng \& Schilling, 1996; Song \& Lee, 2005), Markov Chain Monte Carlo (e.g., Béguin \& Glas, 2001; Edwards, 2010), and stochastic approximation (SA; e.g., Cai, 2010a; Cai, 2010b; Zhang, Chen, \& Liu, 2020). The Metropolis-Hastings Robbins-Monro (MH-RM; Cai, 2010a; 2010b) SA algorithm has been particularly widely used in psychology and in education due to its computational efficiency, and the recent stochastic expectation-maximization (stEM; Zhang, Chen, \& Liu, 2020) algorithm performs comparably to MH-RM. However, even these state-of-the-art SA algorithms are computationally intensive when the sample size and the number of latent factors are very large (e.g., more than $10$ latent factors and more than $\num{10000}$ respondents).

Other marginal likelihood-based parameter estimation methods for exploratory IFA avoid approximating high-dimensional integrals and are therefore more computationally efficient. Limited-information approaches such as the bivariate composite likelihood estimator (Jöreskog \& Moustaki, 2001) and the weighted least squares estimator (Muthén, 1978; 1984) yield fast, consistent, and asymptotically normally distributed estimates but are not asymptotically efficient. Approaches based on variational inference (VI; Jordan et al., 1998; Wainwright \& Jordan, 2008) perform approximate MML estimation by optimizing a lower bound on the marginal likelihood rather than the marginal likelihood itself. More specifically, VI posits a family of approximate latent variable (LV) posterior distributions, then finds the member of this family that is closest to the true LV posterior distribution by optimizing the aforementioned lower bound; the variational estimator is equivalent to the MML estimator when the approximate and true LV posterior distributions are precisely equal. Since a separate set of approximate LV posterior distribution parameters is estimated for each data point, VI's computational complexity depends on the sample size and on the complexity of the approximating family. Variational methods for IFA have demonstrated competitive performance with SA algorithms such as MH-RM for small sample sizes (Cho, 2020; Hui et al., 2017). Additionally, concurrent work by Cho (2020) has established consistency of the variational estimator for the multidimensional two-parameter logistic (M2PL) model in the double asymptotic setting where both the sample size and the number of items simultaneously tend to infinity. However, the variational estimator's theoretical properties have not yet been established for other IFA models or in the classical asymptotic setting where only the sample size tends to infinity.

The MML estimator's computational inefficiency arises from treating the latent factors as random effects that must be integrated out of the marginal likelihood. An alternative class of computationally efficient estimators treats the latent factors as fixed parameters, thereby avoiding the need for specifying a prior distribution on the latent factors and for evaluating high-dimensional integrals. However, these estimators pay a price for their computational efficiency: namely, they are only consistent in the double asymptotic setting described above. The constrained joint maximum likelihood estimator (CJMLE; Chen, Li, \& Zhang, 2019) is the state-of-the-art estimator in this class; it is faster than MML-based approaches and is efficient in the double asymptotic setting. Zhang, Chen, and Li's (2020) estimator based on singular value decomposition (SVD) is faster than CJMLE and does not suffer from convergence issues, although it is not (double) asymptotically efficient.

It is clear that an estimation procedure combining the asymptotic properties of the MML estimator with the computational efficiency of CJMLE is lacking from the IFA literature. In this work, we investigate a VI-based procedure that offers a step toward achieving these properties. This procedure employs techniques from two active deep learning (DL) research areas: amortized variational inference (AVI; Gershman \& Goodman, 2014) and importance weighted variational inference (IWVI; Burda et al., 2016; Domke \& Sheldon, 2018). AVI modifies traditional VI by using a powerful function approximator called an inference model to predict the parameters of the LV posterior for each data point rather than estimating these parameters directly. AVI is faster than traditional VI for large data sets, although it can be less flexible in practice (Cremer et al., 2018). IWVI decreases the gap between the variational lower bound and the true marginal likelihood by drawing multiple importance-weighted (IW) samples from the approximate LV posterior during fitting, thereby trading computational efficiency for a better lower bound. When the number of IW samples equals infinity, IWVI is theoretically equivalent to MML estimation and thus inherits the asymptotic properties of the MML estimator (Burda et al., 2016).

The proposed algorithm is based on the importance-weighted autoencoder (IWAE; Burda et al., 2016), an algorithm for amortized IWVI whose inference model is a deep artificial neural network (ANN). The IWAE is in turn an extension of the variational autoencoder (VAE; Kingma \& Welling, 2014; Rezende et al., 2014), a foundational AVI algorithm that also employs an ANN inference model. Our work extends that of Curi et al. (2019), who used a VAE to estimate item parameters in a confirmatory M2PL model. Our work is also related to concurrent work by Wu et al. (2020), who applied a VAE for confirmatory M2PL item parameter estimation in the Bayesian setting. Our major contributions are as follows: (1) We introduce the IWAE to the IFA literature and describe how it may be applied for exploratory analysis of polytomous item response data in the frequentist setting, and (2) we conduct simulation studies to investigate the finite sample behavior the IWAE and to compare the IWAE to MH-RM and CJMLE.

Our paper is organized as follows. Section \ref{section:2} provides a brief overview of ANNs. Section \ref{section:3} introduces the problem of fitting IFA models with polytomous responses. Section \ref{section:4} describes AVI and IWVI for IFA. The full algorithm is proposed in Section \ref{section:5} and computational details are discussed. Section \ref{section:6} includes an empirical example and simulation studies. Extensions of the method are described in Section \ref{section:7} and discussions are given in Section \ref{section:8}.

\section{A Brief Overview of Artificial Neural Networks} \label{section:2}

Deep learning (DL) models are machine learning models that map a set of predictor variables through a sequence of transformations called layers to predict a set of outcome variables. Much of DL's success in recent years can be attributed to a family of nonlinear statistical models called artificial neural networks (ANNs; LeCun et al., 2015). ANNs are essential building blocks for the algorithm described in this work.

\subsection{Feedforward Neural Networks}

\begin{figure}
    \centering
    \resizebox{0.4\textwidth}{!}{
    \begin{tikzpicture}[]
        \def\nodedist{50pt}
        \def\pindist{20pt}
        \def\layerdist{100pt}
        \def\nodesize{40pt}
        \def\annotpos{20pt}
        
        \tikzstyle{every pin edge}=[signal]
        \tikzstyle{annot} = [text width=4em, text centered]
        
        \foreach \y in {1,...,3}
            \node[inputnode, text width=\nodesize, minimum size=\nodesize]
                (x\y) at (\y*\nodedist,0) {$x_{\y}$};
        
        \node[hiddennode, text width=\nodesize, minimum size=\nodesize] 
            (h11) at ($(x1) + (-0.5*\nodedist, \layerdist)$) {$h_{1}^{(1)}$};
        \foreach \y in {1,...,3}
            {\pgfmathtruncatemacro{\label}{\y + 1}
            \node[hiddennode, text width=\nodesize, minimum size=\nodesize] 
                (h1\label) at ($(x\y) + (0.5*\nodedist, \layerdist)$) {$h_{\label}^{(1)}$};}
        
        \foreach \y in {1,...,2}
            \node[outputnode, text width=\nodesize, minimum size=\nodesize]
                (y\y) at ($(x\y) + (0.5*\nodedist, 2*\layerdist)$) {$y_{\y}$};
           
        \foreach \dest in {1,...,4}
            \foreach \source in {1,...,3}
                \draw[signal] (x\source) -- (h1\dest);
                
        \foreach \dest in {1,...,2}
            \foreach \source in {1,...,4}
                \draw[signal] (h1\source) -- (y\dest);
        
        \node[annot, left=4pt of x1] (x1_lab) {Input layer};
        \node[annot, left=4pt of h11] (h11_lab) {Hidden layer};
        \node[annot, left=4pt of y1] (y1_lab) {Output layer};
            
    \end{tikzpicture}
    }
    \caption{Schematic representation of a feedforward neural network with a single hidden layer. The input layer is a $3 \times 1$ vector, the hidden layer is a $4 \times 1$ vector, and the output layer is a $2 \times 1$ vector. Case subscripts $i$ are omitted to avoid clutter.} \label{fig:1}
\end{figure}
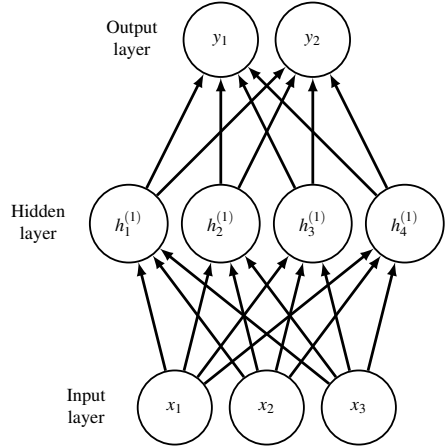

Feedforward neural networks (FNNs) are a simple class of ANNs. In practice, they are used as powerful function approximators because they can approximate any Borel measurable function between finite dimensional spaces to any desired degree of accuracy (Cybenko, 1989). Consider a data set $\{\mathbf{y}_i, \mathbf{x}_i \}_{i = 1}^N$ where $\mathbf{x}_i$ is the $i^\mathrm{th}$ observed $J \times 1$ vector of predictor variables and $\mathbf{y}_i$ is the $i^\mathrm{th}$ observed $P \times 1$ vector of outcome variables. Note that here we define $\mathbf{x}_i$ as a vector of observed variables in line with typical treatments of FNNs, although we will redefine it as a vector of LVs in Section \ref{section:3}. FNNs map the predictor variables through a sequence of $L$ transformations to predict the outcome variables as follows:
\begin{equation} \label{eq:1}
    \mathbf{h}_i^{(l)} = f^{(l)} (\mathbf{W}^{(l)} \mathbf{h}_i^{(l - 1)} + \mathbf{b}^{(l)}), \quad l = 1, \ldots, L,
\end{equation}
where $\mathbf{h}_i^{(0)} = \mathbf{x}_i$, $\mathbf{h}_i^{(L)} = \mathbf{y}_i - \boldsymbol{\varepsilon}_i$ where $\boldsymbol{\varepsilon}_i$ is the $i^\mathrm{th}$ $J \times 1$ vector of errors, $\mathbf{h}_i^{(l)}$ is a $P_l \times 1$ vector of LVs for layers $l = 2, \ldots, L - 1$, $\mathbf{W}^{(l)}$ is a $P_l \times P_{l - 1}$ matrix of regression weights for layer $l$, $\mathbf{b}^{(l)}$ is a $P_{l} \times 1$ vector of intercepts for layer $l$, and $f^{(l)}$ is an almost everywhere differentiable activation function for layer $l$. $\mathbf{x}_i$ is called the input layer, $\mathbf{h}_i^{(1)}, \ldots, \mathbf{h}_i^{(L - 1)}$ are called hidden layers, and $\mathbf{y}_i$ is called the output layer. Figure~\ref{fig:1} shows an FNN schematic diagram.

Notice that FNNs are recursive generalized linear models where each activation function $f^{(l)}$ is an inverse link function linking a linear combination of the variables at layer $l - 1$ to the mean of the variables at layer $l$. In this work, we set the hidden layer activation functions $f^{(1)}, \ldots, f^{(L - 1)}$ to the exponential linear unit (ELU) function
\begin{equation} \label{eq:2}
    f(z) =
        \begin{cases}
            z, & \mathrm{if}\; z \geq 0 \\
            \gamma \big( \exp (z) - 1 \big), & \mathrm{if}\; z < 0
        \end{cases}, \quad z \in \mathbb{R},
\end{equation}
where $\gamma \in \mathbb{R}$ is a hyperparameter (we set $\gamma = 1$ in this work) and $f$ is applied to vectors element-wise. FNNs with ELU hidden layer activation functions are easy to fit and perform well in practice (Clevert et al., 2016). We set the final activation function $f^{(L)}$ to the identity function:
\begin{equation} \label{eq:3}
    f(z) = z, \quad z \in \mathbb{R},
\end{equation}
which is applied to vectors element-wise and corresponds to a linear regression of the layer $L - 1$ LVs on the outcomes.

\subsection{Fitting FNNs Using AMSGrad} \label{section:2.2}

FNNs are typically fitted using stochastic gradient (SG) methods, a class of algorithms that iteratively update model parameters using stochastic estimates of the gradient of the objective function. Readers are referred to Bottou et al. (2018) for an overview of SG methods. In this work, we use the AMSGrad SG algorithm (Reddi et al., 2018), a method that adapts the magnitudes of its parameter updates using exponential moving averages of past stochastic gradient estimates. This approach allows AMSGrad to dynamically utilize information from the observed data to update each parameter a different amount at each iteration (Duchi et al., 2011; McMahan \& Streeter, 2010). AMSGrad has theoretical convergence guarantees and performs well in practice with little tuning. In contrast, approaches based on the Robbins-Monro SA algorithm require the user to pre-specify a sequence of parameter update magnitudes that are fixed across parameters at each fitting iteration. These pre-specified update schemes typically require fine-tuning to the observed data and are often unstable on implementation (Nemirovski et al., 2009; Spall, 2003).

Let $\boldsymbol{\xi}_t$ be a $d \times 1$ vector of parameter values at fitting iteration $t$, $t = 0, \ldots, T$, and let $\mathcal{J}: \mathbb{R}^{d} \mapsto \mathbb{R}$ be a possibly non-convex objective function that decomposes as a sum over observations:
\begin{equation} \label{eq:4}
    \mathcal{J}(\boldsymbol{\xi}_t) = \frac{1}{N} \sum_{i = 1}^N \mathcal{J}_i (\boldsymbol{\xi}_t),
\end{equation}
where $\mathcal{J}_i$ is a per-observation objective fuction. Let $\{\mathbf{y}_i, \mathbf{x}_i \}_{i = 1}^M$ where $M < N$ be a subsample of observations called a mini-batch. Then at iteration $t$, an unbiased estimator of the gradient of the objective function for the full data set is
\begin{equation} \label{eq:5}
    \mathbf{g}_t = \frac{1}{M} \nabla_{\boldsymbol{\xi}_t} \sum_{i = 1}^M \mathcal{J}_i (\boldsymbol{\xi}_t),
\end{equation}
where $\nabla_{\boldsymbol{\xi}_t}$ returns a $d \times 1$ vector of first-order partial derivatives w.r.t. $\boldsymbol{\xi}_t$. AMSGrad proposes iterative parameter updates as follows:
\begin{equation} \label{eq:6}
    \boldsymbol{\xi}_{t + 1} = \boldsymbol{\xi}_t - \eta \frac{\mathbf{m}_t}{\sqrt{\hat{\mathbf{v}}_t}},
\end{equation}
where
\begin{equation} \label{eq:7}
    \begin{split}
        \mathbf{m}_t &= \beta_1 \mathbf{m}_{t - 1} + (1 - \beta_1) \mathbf{g}_t; \\
        \mathbf{v}_t &= \beta_2 \mathbf{v}_{t - 1} + (1 - \beta_2) \mathbf{g}_t^2; \\
        \hat{\mathbf{v}}_t &= \max(\hat{\mathbf{v}}_{t - 1}, \mathbf{v}_t);
    \end{split}
\end{equation}
$\mathbf{m}_0 = \mathbf{0}$; $\mathbf{v}_0 = \mathbf{0}$; $\hat{\mathbf{v}}_0 = \mathbf{0}$; $\mathbf{m}_t$ and $\mathbf{v}_t$ are $d \times 1$ vectors containing exponential moving averages of the gradient and the squared gradient at iteration $t$, respectively; $\beta_1 \in \lbrack 0, 1 \rbrack$ and $\beta_2 \in \lbrack 0, 1 \rbrack$ are forgetting factors for the gradient and the squared gradient, respectively; $\eta > 0$ is a step size called the learning rate; and square, square root, division, and maximum operations are applied to vectors element-wise. When mini-batches are sampled uniformly at random with replacement and the learning rate is sufficiently small, AMSGrad is guaranteed to converge to a local stationary point for smooth, non-convex objective functions. Zhou et al. (2018) as well as Chen, Liu, Sun, and Hong (2019) provide conditions required for first-order convergence, while Staib et al. (2019) discuss second-order convergence. Importantly, computation time per iteration does not increase with the sample size, allowing for convergence even with very large-scale data (Bottou et al., 2018).

When $\boldsymbol{\xi}_t$ are FNN parameters, an algorithm called backpropagation (BP) is used to efficiently compute the gradient estimator in equation~\ref{eq:5}. BP is an application of the chain rule of calculus and is a special case of reverse mode automatic differentiation (Linnainmaa, 1970). Goodfellow et al. (2016) provide a detailed discussion of BP.

\section{The Problem of Fitting High-Dimensional Item Factor Analysis Models} \label{section:3}

\subsection{The Graded Item Response Model}

Samejima's (1969) graded response model (GRM) is a widespread model for polytomous item responses. We introduce notation for the GRM following Cai (2010a). Suppose there are $i = 1, \ldots, N$ distinct respondents and $j = 1, \ldots, J$ items. Let $y_{i,j} \in \{0, 1, \ldots, C_j - 1\}$ denote the response for respondent $i$ to item $j$ in $C_j$ graded (i.e., ordinal) categories. Note that when $C_j = 2$ for all $j$, the GRM reduces to the M2PL (McKinley \& Reckase, 1983).

Suppose we have $P$ LVs; let $\mathbf{x}_i$ denote the $P \times 1$ vector of factor scores (i.e., LV values) for respondent $i$. Let $\boldsymbol{\beta}_j$ denote the $P \times 1$ vector of loadings, let $\boldsymbol{\alpha}_j = (\alpha_{j1}, \ldots, \alpha_{j, C_j - 1})^\top$ denote the $(C_j - 1) \times 1$ vector of strictly ordered category intercepts, and let $\boldsymbol{\theta}_j = (\boldsymbol{\alpha}_j^\top, \boldsymbol{\beta}_j^\top)^\top$ be a vector collecting all parameters for item $j$. The GRM defines a set of boundary response probabilities conditional on the item parameters $\boldsymbol{\theta}_j$ and the factor scores $\mathbf{x}_i$:
\begin{equation} \label{eq:8}
    \Pr(y_{i,j} \geq k \mid \boldsymbol{\theta}_j, \mathbf{x}_i) =
    \frac{1}{1 + \mathrm{exp} \big[ {-D}(\alpha_{j,k} + \boldsymbol{\beta}_j^\top \mathbf{x}_i) \big]}, \quad k \in \{1, \ldots, C_j - 1\},
\end{equation}
where $\Pr(y_{i,j} \geq 0 \mid \boldsymbol{\theta}_j, \mathbf{x}_i) = 1$, $\Pr(y_{i,j} \geq C_j \mid \boldsymbol{\theta}_j, \mathbf{x}_i) = 0$, and $D$ is a scaling constant (typically $1.702$) used to help the logistic metric better approximate the normal ogive metric (Reckase, 2009). The conditional probability for a particular response $y_{i,j} = k$, $k \in \{0, \ldots, C_j - 1\}$ is
\begin{equation} \label{eq:9}
    \pi_{i,j,k} = P(y_{i,j} = k \mid \boldsymbol{\theta}_j, \mathbf{x}_i) = \Pr(y_{i,j} \geq k \mid \boldsymbol{\theta}_j, \mathbf{x}_i) - \Pr(y_{i,j} \geq k + 1 \mid \boldsymbol{\theta}_j, \mathbf{x}_i).
\end{equation}

\subsection{Observed Data Likelihood}

It follows from equation~\ref{eq:9} that the conditional distribution of $y_{i,j}$ is multinomial with $C_j$ cells, trial size $1$, and cell probabilities $\pi_{i,j,k}$:
\begin{equation} \label{eq:10}
    p_{\boldsymbol{\theta}_j}(y_{i,j}\mid \mathbf{x}_i) = \prod_{k = 0}^{C_j - 1} \pi_{i,j,k}^{\mathbbm{1}_k(y_{i,j})},
\end{equation}
where we define the indicator function
\begin{equation} \label{eq:11}
    \mathbbm{1}_k(y) =
        \begin{cases}
            1, & \text{if}\; y = k \\
            0, & \text{otherwise}
        \end{cases}
\end{equation}
for $k \in \{ 0, \ldots C_j - 1\}$. Let $\mathbf{y}_i = (y_{i,1}, \ldots, y_{i,n})^\top$ be respondent $i$'s response pattern. By the usual conditional independence assumption, the conditional distribution of $\mathbf{y}_i$ is
\begin{equation} \label{eq:12}
    p_{\boldsymbol{\theta}}(\mathbf{y}_i \mid \mathbf{x}_i) = \prod_{j = 1}^J p_{\boldsymbol{\theta}_j}(y_{i,j}\mid \mathbf{x}_i),
\end{equation}
where $\boldsymbol{\theta}$ is a vector collecting the estimable parameters for all $J$ items.

Assume the prior distribution of factor scores $\mathbf{x}_i$ is standard multivariate normal with density function $\mathcal{N}( \mathbf{x}_i)$. Then the marginal distribution of $\mathbf{y}_i$ is given by
\begin{equation} \label{eq:13}
    p_{\boldsymbol{\theta}}(\mathbf{y}_i) = \int \prod_{j = 1}^J p_{\boldsymbol{\theta}_j}(y_{i,j}\mid \mathbf{x}) \mathcal{N}( \mathbf{x})d\mathbf{x},
\end{equation}
where the above integral is over $\mathbb{R}^p$. Let $\mathbf{Y}$ be an $N \times J$ matrix of independent response patterns whose $i^\mathrm{th}$ row is $\mathbf{y}_i^\top$. The observed data likelihood is
\begin{equation} \label{eq:14}
    \mathcal{L}(\boldsymbol{\theta} \mid \mathbf{Y}) = \prod_{i = 1}^N \bigg[ \int \prod_{j = 1}^J p_{\boldsymbol{\theta}_j}(y_{i,j}\mid \mathbf{x}) \mathcal{N}( \mathbf{x})d\mathbf{x} \bigg].
\end{equation}
Maximizing $\mathcal{L}(\boldsymbol{\theta} \mid \mathbf{Y})$ directly is difficult because we must approximate the $N$ integrals over $\mathbb{R}^p$ numerically. In this work, we avoid this difficulty by deriving an analytical lower bound on $\log \mathcal{L}(\boldsymbol{\theta} \mid \mathbf{Y})$ using VI. We then maximize this lower bound using a DL algorithm.

\section{Variational Methods for Item Factor Analysis} \label{section:4}

Variational inference (VI) is an approach to approximate maximum likelihood estimation for LV models that has recently gained traction in the machine learning community (Blei et al., 2017; Zhang, Butepage, Kjellstrom, \& Mandt, 2019). VI has been applied for confirmatory IFA in both the frequentist setting (Cho, 2020; Curi et al., 2019) and the Bayesian setting (Chen, Filho, Prudêncio, Diethe, \& Flach, 2019; Natesan et al., 2016; Wu et al., 2020) as well as for exploratory IFA in the frequentist setting (Cho, 2020; Hui et al., 2017). In this section, we describe variational methods for IFA as well as an importance sampling technique for helping the variational estimator better approximate the MML estimator.

\subsection{Variational Inference}

We now describe VI in the context of a general LV model, then apply VI to IFA in the following sections. The main idea behind VI is to treat LV inference as an optimization problem. More formally, let $\mathbf{y} \in \mathcal{Y}$ and $\mathbf{x} \in \mathcal{X}$ denote observed and LVs, respectively, where $\mathcal{Y}$ and $\mathcal{X}$ are sample spaces. VI introduces a family $\mathcal{Q}$ of approximate densities over LVs and aims to find the member $q_{\boldsymbol{\psi}^*(\mathbf{y})}(\mathbf{x}) \in \mathcal{Q}$ that minimizes the Kullback-Leibler (KL) divergence\footnote{For distributions $q$ and $p$, the KL divergence is defined as $\dkl \big[ q \| p \big] = \mathbb{E}_{q} \big[ \log q \big] - \mathbb{E}_{q} \big[ \log p \big] $. It can be shown that $\dkl \big[ q \| p \big] \geq 0$ with equality if and only if $p = q$ almost everywhere w.r.t. $q$.} from itself to the true LV posterior:
\begin{equation} \label{eq:15}
    q_{\boldsymbol{\psi}^*(\mathbf{y})}(\mathbf{x}) = \argmin_{q_{\boldsymbol{\psi}(\mathbf{y})}(\mathbf{x}) \in \mathcal{Q}} \dkl \big[ q_{\boldsymbol{\psi}(\mathbf{y})}(\mathbf{x}) \| p (\mathbf{x} \mid \mathbf{y}) \big],
\end{equation}
where $\boldsymbol{\psi}(\mathbf{y})$ is a vector of variational parameters. Note that $\boldsymbol{\psi}(\mathbf{y})$ depends on $\mathbf{y}$, indicating that a different vector of variational parameters is estimated for each observation. For models with continuous LVs, an often tractable choice for the approximate posterior is the isotropic normal density (Kingma \& Welling, 2014):
\begin{equation} \label{eq:16}
    q_{\boldsymbol{\psi}(\mathbf{y})}(\mathbf{x}) = \mathcal{N}\big(\mathbf{x} \mid \boldsymbol{\mu}(\mathbf{y}), \boldsymbol{\sigma}^2( \mathbf{y})\mathbf{I}_P \big), 
\end{equation}
where $\boldsymbol{\mu}(\mathbf{y})$ is a $P \times 1$ vector of means, $\boldsymbol{\sigma}^2(\mathbf{y})$ is a $P \times 1$ vector of variances, and $\mathbf{I}_P$ is a $P \times P$ identity matrix. Minimizing the KL divergence from the isotropic normal approximate posterior to the true LV posterior produces the ``best'' isotropic normal approximation to the true LV posterior. In practice, however, tractable approximate posteriors such as the isotropic normal density are rarely flexible enough to perfectly approximate the true LV posterior and thereby minimize the KL divergence to zero. The importance sampling technique described later in this section improves the accuracy of VI by implicitly increasing the flexibility of the approximate posterior.

\subsection{Evidence Lower Bound}

The log-likelihood of the observed data under the GRM can be written as a sum over the marginal likelihood of each observation:
\begin{equation}
    \ell(\boldsymbol{\theta} \mid \mathbf{Y}) = \sum_{i = 1}^N \log p_{\boldsymbol{\theta}}(\mathbf{y}_i) \label{eq:17},
\end{equation}
where $\ell(\boldsymbol{\theta} \mid \mathbf{Y}) = \log \mathcal{L}(\boldsymbol{\theta} \mid \mathbf{Y})$. Let the approximate LV posterior be the isotropic normal density as in equation~\ref{eq:16}. We can re-write a single summand in equation~\ref{eq:17} as
\begin{equation}
    \log p_{\boldsymbol{\theta}}(\mathbf{y}_i) = \dkl \big[ q_{\boldsymbol{\psi}(\mathbf{y}_i)}(\mathbf{x}_i) \| p_{\boldsymbol{\theta}}(\mathbf{x}_i \mid \mathbf{y}_i) \big] + \mathbb{E}_{q_{\boldsymbol{\psi}(\mathbf{y}_i)}(\mathbf{x}_i)}\big[ \log p_{\boldsymbol{\theta}}(\mathbf{x}_i, \mathbf{y}_i) - \log q_{\boldsymbol{\psi}(\mathbf{y}_i)}(\mathbf{x}_i) \big]. \label{eq:18}
\end{equation}
The first term on the r.h.s. of equation~\ref{eq:18} is the KL divergence from the approximate to the true LV posterior (i.e., it is the term we wish to minimize from equation~\ref{eq:15}). Since this term is non-negative, the second term on the r.h.s. of equation~\ref{eq:18} is a lower bound on the marginal likelihood of a single observation. This term is called the evidence lower bound (ELBO) and can be re-written as
\begin{align}
    \log p_{\boldsymbol{\theta}}(\mathbf{y}_i) &\geq \mathbb{E}_{q_{\boldsymbol{\psi}(\mathbf{y}_i)}(\mathbf{x}_i)}\big[ \log p_{\boldsymbol{\theta}}(\mathbf{x}_i, \mathbf{y}_i) - \log q_{\boldsymbol{\psi}(\mathbf{y}_i)}(\mathbf{x}_i) \big] \label{eq:19} \\
    &= \mathbb{E}_{q_{\boldsymbol{\psi}(\mathbf{y}_i)}(\mathbf{x}_i)}\big[ \log p_{\boldsymbol{\theta}}(\mathbf{y}_i \mid \mathbf{x}_i) \big] - \dkl \big[q_{\boldsymbol{\psi}(\mathbf{y}_i)}(\mathbf{x}_i) \| p_{\boldsymbol{\theta}}(\mathbf{x}_i)\big] \label{eq:20} \\
    &= \elbo_i \label{eq:21}.
\end{align}
The first term in the ELBO on line~\ref{eq:20} is an expected conditional log-likelihood that encourages ${q^*(\mathbf{x}_i \mid \mathbf{y}_i)}$ to place mass on LVs that explain the observed data well, while the second term encourages densities that are close to the LV prior $p_{\boldsymbol{\theta}}(\mathbf{x}_i)$. Maximizing the ELBO over all observations w.r.t. the item parameters $\boldsymbol{\theta}$ and the variational parameters $\boldsymbol{\psi}(\mathbf{y}_i)$ both approximately maximizes the observed data log-likelihood and minimizes the KL divergence from the approximate to the true LV posterior.

\subsection{Amortized Variational Inference}

Traditional VI fits a different approximate LV posterior for each observation, which quickly becomes computationally infeasible for large data sets. It is also not straightforward to apply models fitted using VI to previously unseen observations (e.g., to perform LV inference for or to compute the log-likelihood of the unseen observations). Amortized variational inference (AVI) is a computationally efficient alternative to VI that uses a powerful function approximator called an inference model to parameterize the approximate posterior. By sharing the parameters of the inference model across observations, AVI estimates a constant number of parameters regardless of the sample size, whereas VI estimates a number of parameters that at best grows linearly as a function of the sample size. Further, models fitted using AVI can easily be applied to previously unseen observations by simply feeding the observations to the inference model.

The variational autoencoder (VAE; Kingma \& Welling, 2014; Rezende et al., 2014) is an AVI algorithm whose inference model is an ANN. We can use a VAE for IFA by parameterizing the approximate LV posterior as follows:
\begin{align} \label{eq:22}
    \begin{split}
        \big( \boldsymbol{\mu}_i^\top, \log \boldsymbol{\sigma}_i^\top \big)^\top &= \fnn_{\boldsymbol{\psi}}(\mathbf{y}_i), \\
        q_{\boldsymbol{\psi}}(\mathbf{x}_i \mid \mathbf{y}_i) &= \mathcal{N}\big(\mathbf{x}_i \mid \boldsymbol{\mu}_i, \boldsymbol{\sigma}^2_i\mathbf{I}_P \big),
    \end{split}
\end{align}
where $\boldsymbol{\mu}_i$ is a $P \times 1$ predicted vector of means, $\log \boldsymbol{\sigma}_i$ is a $P \times 1$ predicted vector of log-standard deviations, and $\fnn_{\boldsymbol{\psi}}$ is an $L$-layer FNN parameterized by $\boldsymbol{\psi}$. Rather than estimating a set of variational parameters $\boldsymbol{\psi}(\mathbf{y}_i)$ for each observation, the FNN parameters $\boldsymbol{\psi}$ are now shared across observations. That is, rather than maximizing equation~\ref{eq:20} over observations, we now maximize
\begin{equation}
    \elbo = \mathbb{E}_{q_{\boldsymbol{\psi}}(\mathbf{x} \mid \mathbf{y})}\big[ \log p_{\boldsymbol{\theta}}(\mathbf{y} \mid \mathbf{x}) \big] - \dkl \big[q_{\boldsymbol{\psi}}(\mathbf{x} \mid \mathbf{y}) \| p_{\boldsymbol{\theta}}(\mathbf{x})\big] \label{eq:23}
\end{equation}
over observations. Note that we now drop the case index $i$ since the FNN parameters $\boldsymbol{\psi}$ are shared across $\{ \mathbf{y}_i, \mathbf{x}_i \}_{i = 1}^N$. In theory, the VAE is equivalent to VI when the FNN is sufficiently flexible (e.g., when the FNN has one infinitely large hidden layer). In practice, the FNN has finite capacity and may prevent the VAE from performing as well as VI. This performance difference is called the amortization gap and may be reduced by increasing the flexibility of the approximate LV posterior (Cremer et al., 2018).

\subsection{Fitting the Amortized Model}

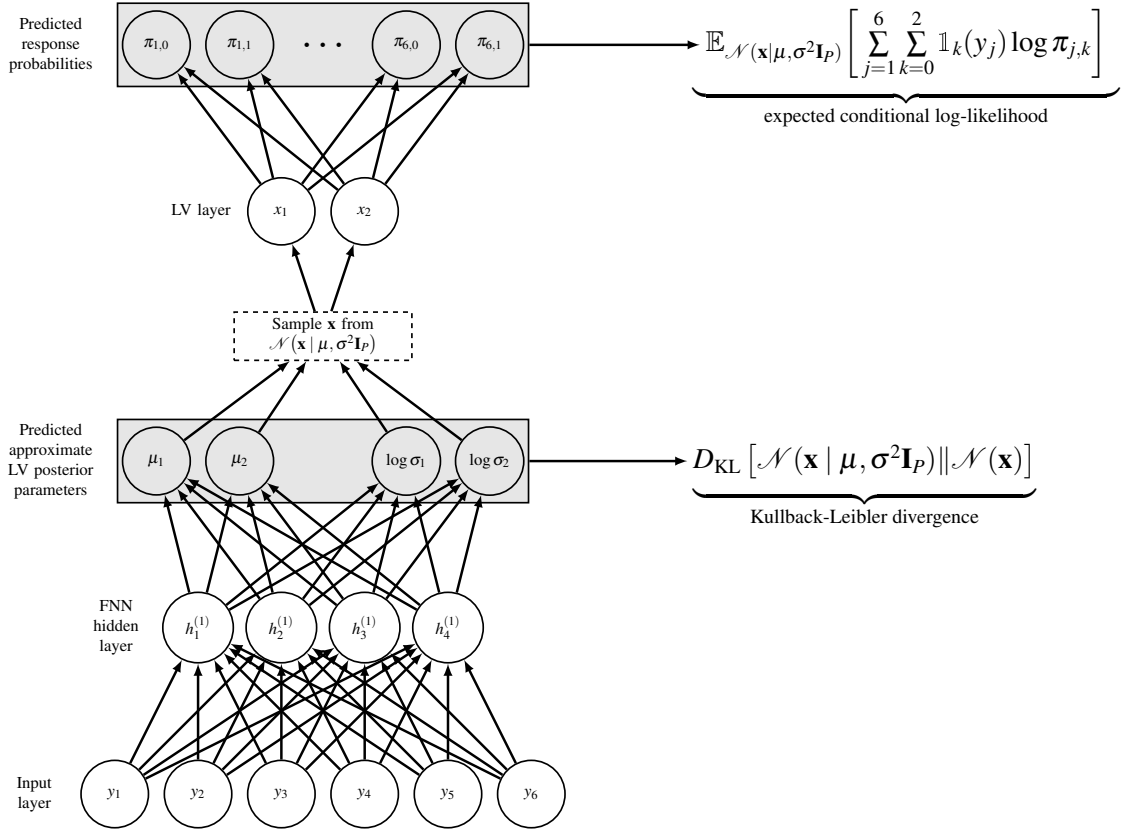
\begin{figure}[ht]
	\centering
    \resizebox{\textwidth}{!}{
	    \begin{tikzpicture}[]
	        \def\nodedist{50pt}
            \def\pindist{20pt}
	        \def\layerdist{100pt}
            \def\nodesize{40pt}
            \def\annotpos{20pt}
            
            \tikzstyle{every pin edge}=[signal]
            \tikzstyle{annot} = [text width=4em, text centered]
            
            \foreach \y in {1,...,6}
                \node[inputnode, text width=\nodesize, minimum size=\nodesize] 
                    (y\y) at (\y*\nodedist,0) {$y_{\y}$};
            
            \foreach \y in {1,...,4}
                \node[hiddennode, text width=\nodesize, minimum size=\nodesize] 
                    (h1\y) at ($(y\y) + (\nodedist, \layerdist)$) {$h_{\y}^{(1)}$};
            
            \foreach \y in {1,...,2}
                \node[inputnode, text width=\nodesize, minimum size=\nodesize]
                    (mu\y) at ($(h1\y) + (-0.5*\nodedist, \layerdist)$) {$\mu_{\y}$};
                    
            \foreach \y in {1,...,2}
                \node[inputnode, text width=\nodesize, minimum size=\nodesize]
                    (sigma\y) at ($(h1\y) + (2.5*\nodedist, \layerdist)$) {$\log \sigma_{\y}$};
                    
            \node[fill=white, dashed, draw=black, text width=100pt, align=center, line width=1.0pt]
                (normal) at ($(mu1) + (2*\nodedist, 0.75*\layerdist)$) {Sample $\mathbf{x}$ from ${\mathcal{N}\big(\mathbf{x} \mid \boldsymbol{\mu}, \boldsymbol{\sigma}^2 \mathbf{I}_P \big)}$};
            
            \foreach \y in {1,...,2}
                \node[inputnode, text width=\nodesize, minimum size=\nodesize]
                    (x\y) at ($(y\y) + (2*\nodedist, 3.5*\layerdist)$) {$x_{\y}$};
            \foreach \y in {1,...,2}
                {\pgfmathtruncatemacro{\category}{\y - 1}
                \node[outputnode, text width=\nodesize, minimum size=\nodesize]
                    (pi1\category) at ($(mu\y) + (0, 2.5*\layerdist)$) {$\pi_{1,\category}$};}
            \foreach \y in {1,...,2}
                {\pgfmathtruncatemacro{\category}{\y - 1}
                \node[outputnode, text width=\nodesize, minimum size=\nodesize]
                    (pi6\category) at ($(sigma\y) + (0, 2.5*\layerdist)$) {$\pi_{6,\category}$};}
            \node[annot] (pidots) at ($(y3) + (0.5*\nodedist, 4.5*\layerdist)$) {\Huge $\ldots$};
                    
            \node[draw=black, fill=black, fill opacity=0.1, text width=240pt, align=center, minimum height = \nodedist, line width=1.0pt]
                (paramrect1) at ($(normal) - (0, 0.75*\layerdist)$) {};
                
            \node[draw=black, fill=black, fill opacity=0.1, text width=240pt, align=center, minimum height = \nodedist, line width=1.0pt]
                (paramrect2) at ($(normal) + (0, 1.75*\layerdist)$) {};
            
            \node[align=center]
                (kld) at ($(paramrect1) + (3.25*\layerdist, 0)$) {\LARGE $\dkl \big[\mathcal{N}(\mathbf{x} \mid \boldsymbol{\mu}, \boldsymbol{\sigma}^2 \mathbf{I}_P) \| \mathcal{N}( \mathbf{x} )\big]$};
                
            \node[align=center]
                (underbrace1) at ($(kld) - (0, 0.35*\nodedist)$) {\LARGE $\underbrace{\phantom{\dkl \big[\mathcal{N}(\mathbf{x} \mid \boldsymbol{\mu}, \boldsymbol{\sigma}^2 \mathbf{I}_P) \| \mathcal{N}( \mathbf{x} )\big]}}_{\text{Kullback-Leibler divergence}}$};
            
            \node[align=center]
                (likelihood) at ($(kld) + (0.5*\nodedist, 2.5*\layerdist)$) {\LARGE $\mathbb{E}_{\mathcal{N}(\mathbf{x} \mid \boldsymbol{\mu}, \boldsymbol{\sigma}^2 \mathbf{I}_P)} \bigg[ \sum\limits_{j = 1}^6 \sum\limits_{k = 0}^{2} \mathbbm{1}_k(y_{j}) \log \pi_{j,k} \bigg]$};
                
            \node[align=center]
                (underbrace2) at ($(likelihood) - (0, 0.25*\nodedist)$) {\LARGE $\underbrace{\phantom{\mathbb{E}_{\mathbf{x} \sim \mathcal{N}(\mathbf{x} \mid \boldsymbol{\mu}, \boldsymbol{\sigma}^2 \mathbf{I}_P)} \bigg[ \sum\limits_{j = 1}^6 \sum\limits_{k = 0}^{2} \mathbbm{1}_k(y_{j}) \log \pi_{j,k} \bigg]}}_{\text{expected conditional log-likelihood}}$};
               
            \foreach \dest in {1,...,4}
                \foreach \source in {1,...,6}
                    \draw[signal] (y\source) -- (h1\dest);
                    
            \foreach \dest in {1,...,2}
                \foreach \source in {1,...,4}
                    \draw[signal] (h1\source) -- (mu\dest);
                    
            \foreach \dest in {1,...,2}
                \foreach \source in {1,...,4}
                    \draw[signal] (h1\source) -- (sigma\dest);
                    
            \foreach \source in {1,...,2}
                \draw[signal] (mu\source) -- (normal);
                
            \foreach \source in {1,...,2}
                \draw[signal] (sigma\source) -- (normal);
                
            \foreach \dest in {1,...,2}
                \draw[signal] (normal) -- (x\dest);
                    
            \foreach \dest in {0,...,1}
                \foreach \source in {1,...,2}
                    \draw[signal] (x\source) -- (pi1\dest);
            \foreach \dest in {0,...,1}
                \foreach \source in {1,...,2}
                    \draw[signal] (x\source) -- (pi6\dest);
                    
            \draw[signal] (paramrect1) -- (kld);
            
            \draw[signal] (paramrect2) -- (likelihood);
            
            \node[annot, left=4pt of y1] (y1_lab) {Input layer};
            \node[annot, left=4pt of h11] (h1_lab) {FNN hidden layer};
            \node[annot, text width = 70pt, left=4pt of mu1] (mu_lab) {Predicted approximate LV posterior parameters};
            \node[annot, left=4pt of x1] (x1_lab) {LV layer};
            \node[annot, text width = 70pt, left=4pt of pi10] (pi_lab) {Predicted response probabilities};
                
	    \end{tikzpicture}
    }
    \caption{Schematic diagram of a variational autoencoder for item factor analysis with $J = 6$ items, $C_j = 2$ categories per item, $P = 2$ factors, $S = 1$ Monte Carlo sample from the approximate latent variable posterior, and an inference model consisting of a feedforward neural network with a single hidden layer. The reparameterization trick is not illustrated for simplicity. LV = latent variable.} \label{fig:2}
\end{figure}

Fitting the VAE for IFA can be accomplished with AMSGrad and BP after obtaining an unbiased estimator for the gradient of the ELBO w.r.t. the model parameters $\boldsymbol{\theta}$ and $\boldsymbol{\psi}$. An unbiased estimator for the gradient of the ELBO w.r.t. the item parameters $\boldsymbol{\theta}$ is
\begin{align}
    \nabla_{\boldsymbol{\theta}} \elbo &= \nabla_{\boldsymbol{\theta}} \mathbb{E}_{q_{\boldsymbol{\psi}}(\mathbf{x} \mid \mathbf{y})} \big[ \log p_{\boldsymbol{\theta}}(\mathbf{x}, \mathbf{y}) - \log q_{\boldsymbol{\psi}}(\mathbf{x} \mid \mathbf{y}) \big] \label{eq:24} \\
    &= \mathbb{E}_{q_{\boldsymbol{\psi}}(\mathbf{x} \mid \mathbf{y})}\big[ \nabla_{\boldsymbol{\theta}}\log p_{\boldsymbol{\theta}}(\mathbf{x}, \mathbf{y}) \big] \label{eq:25} \\
    &\approx \frac{1}{S} \sum_{s = 1}^S \nabla_{\boldsymbol{\theta}}\log p_{\boldsymbol{\theta}}(\mathbf{y}, \mathbf{x}_s), \label{eq:26}
\end{align}
where line~\ref{eq:26} approximates the expectations in line~\ref{eq:25} with a size $S$ Monte Carlo sample of factor scores from the approximate LV posterior.\footnote{We move the gradient inside the expectation in line~\ref{eq:25} using the fact that $q_{\boldsymbol{\psi}}(\mathbf{x} \mid \mathbf{y})$, $\log q_{\boldsymbol{\psi}}(\mathbf{x} \mid \mathbf{y})$, and $\log p_{\boldsymbol{\theta}}(\mathbf{x}, \mathbf{y})$ satisfy certain regularity conditions. For details, see Lehmann and Casella (1998).} Obtaining an unbiased estimator for the gradient of the ELBO w.r.t. the FNN parameters $\boldsymbol{\psi}$ is more challenging because, in general,
\begin{align}
    \nabla_{\boldsymbol{\psi}} \elbo &= \nabla_{\boldsymbol{\psi}} \mathbb{E}_{q_{\boldsymbol{\psi}}(\mathbf{x} \mid \mathbf{y})} \big[ \log p_{\boldsymbol{\theta}}(\mathbf{x}, \mathbf{y}) - \log q_{\boldsymbol{\psi}}(\mathbf{x} \mid \mathbf{y}) \big] \label{eq:27} \\
    &\neq \mathbb{E}_{q_{\boldsymbol{\psi}}(\mathbf{x} \mid \mathbf{y})} \big[ \nabla_{\boldsymbol{\psi}}\log p_{\boldsymbol{\theta}}(\mathbf{x}, \mathbf{y}) - \nabla_{\boldsymbol{\psi}}\log q_{\boldsymbol{\psi}}(\mathbf{x} \mid \mathbf{y}) \big], \label{eq:28}
\end{align}
since the expectations are taken w.r.t. $q_{\boldsymbol{\psi}} (\mathbf{x} \mid \mathbf{y})$, which is a function of $\boldsymbol{\psi}$. To overcome this problem, we reparameterize $\mathbf{x}$ as follows:
\begin{align} \label{eq:29}
    \begin{split}
    \boldsymbol{\epsilon} &\sim \mathcal{N} (\boldsymbol{\epsilon}), \\
    \mathbf{x} &= \boldsymbol{\mu} + \boldsymbol{\sigma} \odot \boldsymbol{\epsilon},
    \end{split}
\end{align}
where $\boldsymbol{\epsilon}$ is a $P \times 1$ sample from a standard multivariate normal density, $\boldsymbol{\mu}$ and $\boldsymbol{\sigma}$ are the outputs of the FNN inference model given in equations~\ref{eq:22}, and $\odot$ denotes element-wise multiplication. This reparameterization ``externalizes'' the randomness in $\mathbf{x}$ by writing $\mathbf{x}$ as a deterministic function of $\boldsymbol{\psi}$. We can now obtain an unbiased estimator for the gradient of the ELBO w.r.t. $\boldsymbol{\psi}$ as follows:
\begin{align}
    \nabla_{\boldsymbol{\psi}} \elbo &= \nabla_{\boldsymbol{\psi}}\mathbb{E}_{\mathcal{N}(\boldsymbol{\epsilon})} \big[ \log p_{\boldsymbol{\theta}}(\mathbf{x}, \mathbf{y}) - \log q_{\boldsymbol{\psi}}(\mathbf{x} \mid \mathbf{y}) \big] \label{eq:30}  \\
    &= \mathbb{E}_{\mathcal{N} (\boldsymbol{\epsilon})}\big[ \nabla_{\boldsymbol{\psi}}\log p_{\boldsymbol{\theta}}(\mathbf{x}, \mathbf{y}) - \nabla_{\boldsymbol{\psi}}\log q_{\boldsymbol{\psi}}(\mathbf{x} \mid \mathbf{y}) \big] \label{eq:31} \\
    &\approx \frac{1}{S} \sum_{s = 1}^S \big[ \nabla_{\boldsymbol{\psi}}\log p_{\boldsymbol{\theta}}(\mathbf{x}_s, \mathbf{y}) - \nabla_{\boldsymbol{\psi}}\log q_{\boldsymbol{\psi}}(\mathbf{x}_s \mid \mathbf{y}) \big], \label{eq:32}
\end{align}
where the expectations are now taken w.r.t. $\mathcal{N}(\boldsymbol{\epsilon})$ and line~\ref{eq:32} is a Monte Carlo approximation to the expectation in line~\ref{eq:31}. Figure~\ref{fig:2} illustrates how computation proceeds in a VAE for IFA. We note that the KL divergence term shown has a closed form that is efficient to compute (Kingma \& Welling, 2014):
\begin{equation} \label{eq:33}
    \dkl \big[\mathcal{N}(\mathbf{x} \mid \boldsymbol{\mu}, \boldsymbol{\sigma}^2 \mathbf{I}_P) \| \mathcal{N}( \mathbf{x} )\big] = \frac{1}{2} \sum_{p = 1}^P \big[\mu_p^2 + \sigma^2_p - 1 - \log \sigma^2_p \big].
\end{equation}

\subsection{Importance-Weighted Variational Inference}

Importance-weighted variational inference (IWVI; Burda et al., 2016; Domke \& Sheldon, 2018) is a VI strategy that can approximate the true log-likelihood arbitrarily well. Amortized IWVI for IFA maximizes a lower bound called the importance-weighted ELBO (IW-ELBO):
\begin{align}
    \log p_{\boldsymbol{\theta}}(\mathbf{y}) &\geq \iwelbo \label{eq:34} \\
    &= \mathbb{E}_{\mathbf{x}_{1:R}}\bigg[ \log \frac{1}{R} \sum_{r = 1}^R w_r \bigg] \label{eq:35},
\end{align}
where $\mathbf{x}_{1:R} \sim \prod_{r = 1}^R q_{\boldsymbol{\psi}}(\mathbf{x}_r \mid \mathbf{y})$, $w_r = p_{\boldsymbol{\theta}}(\mathbf{x}_r, \mathbf{y}) / q_{\boldsymbol{\psi}}(\mathbf{x}_r \mid \mathbf{y})$ are unnormalized importance weights for the joint distribution of latent and observed variables, and $R$ is the number of importance-weighted (IW) samples. When $R = 1$, the IW-ELBO reduces to the ELBO. As $R \rightarrow \infty$, the IW-ELBO converges monotonically to the marginal log-likelihood (Burda et al., 2016). IWVI also implicitly defines a flexible approximate LV posterior $q_{\boldsymbol{\psi}}^{\text{IW}}(\mathbf{x} \mid \mathbf{y})$ that converges to the true LV posterior pointwise as $R \rightarrow \infty$ (Cremer et al., 2017). These facts imply that IWVI is equivalent to MML estimation when the number of importance samples $R$ equals infinity, in which case IWVI inherits the MML estimator's asymptotic properties. When the inference model is an FNN, the associated IWVI algorithm is called the importance-weighted autoencoder (IWAE; Burda et al., 2016).

Optimizing the IW-ELBO permits trading computational efficiency for a better approximation to the MML estimator by increasing $R$. As with the ELBO, we can obtain an unbiased estimator for the gradient of the IW-ELBO w.r.t. $\boldsymbol{\xi} = (\boldsymbol{\theta}^\top, \boldsymbol{\psi}^\top)^\top$ via the reparameterization trick:
\begin{align}
    \nabla_{\boldsymbol{\xi}} \mathbb{E}_{\mathbf{x}_{1:R}}\bigg[ \log \frac{1}{R} \sum_{r = 1}^R w_r \bigg] &= \mathbb{E}_{\boldsymbol{\epsilon}_{1:R}} \bigg[ \sum_{r = 1}^R \widetilde{w}_r \nabla_{\boldsymbol{\xi}} \log w_r \bigg] \\
    &\approx \frac{1}{S} \sum_{s = 1}^S \bigg[ \sum_{r = 1}^R \widetilde{w}_{r, s} \nabla_{\boldsymbol{\xi}} \log w_{r, s} \bigg],
\end{align}
where $\boldsymbol{\epsilon}_{1:R} \sim \prod_{r = 1}^R\mathcal{N}(\boldsymbol{\epsilon}_r)$ and $\widetilde{w}_r = w_r / \sum_{r' = 1}^R w_{r'}$ are normalized importance weights. Problematically, however, increasing $R$ degrades performance of the above gradient estimator for the inference model parameters $\boldsymbol{\psi}$ (but not for the item parameters $\boldsymbol{\theta}$). Specifically, Rainforth et al. (2018) theoretically and empirically show that as $R$ increases, the signal-to-noise ratio (SNR) of the inference model gradient estimator tends to zero so that the estimator becomes completely random. We resolve this issue using Tucker et al.'s (2019) doubly reparameterized gradient (DReG) estimator:
\begin{align}
    \nabla_{\boldsymbol{\psi}} \mathbb{E}_{\mathbf{x}_{1:R}}\bigg[ \log \frac{1}{R} \sum_{r = 1}^R w_r \bigg] &= \mathbb{E}_{\boldsymbol{\epsilon}_{1:R}}\bigg[ \sum_{r = 1}^R \widetilde{w}_r^2 \frac{\partial \log w_r}{\partial \mathbf{x}_r} \frac{\partial \mathbf{x}_r}{\partial \boldsymbol{\psi}} \bigg] \label{eq:4.21} \\
    &\approx \frac{1}{S} \sum_{s = 1}^S \bigg[ \sum_{r = 1}^R \widetilde{w}_{r, s}^2 \frac{\partial \log w_{r, s}}{\partial \mathbf{x}_{r, s}} \frac{\partial \mathbf{x}_{r, s}}{\partial \boldsymbol{\psi}} \bigg] \label{eq:4.22}.
\end{align}
The DReG estimator is unbiased, has increasing SNR as $R \rightarrow \infty$, and empirically demonstrates lower variance than alternative estimators. In practice, the IW-ELBO $\boldsymbol{\theta}$-gradient and DReG estimators can be successfully approximated using a single Monte Carlo sample (e.g., Burda et al., 2016; Tucker et al., 2019), so we set $S = 1$ for all numerical examples in this work.

\section{Implementation Details} \label{section:5}

\subsection{Starting Values}

The proposed algorithm is detailed in Algorithm~\ref{algorithm:1}. We now discuss choosing the algorithm starting values $\boldsymbol{\xi}_0 = (\boldsymbol{\theta}^\top_0, \boldsymbol{\psi}^\top_0)^\top$.

\begin{algorithm}
\caption{\textit{Deep Learning Algorithm for Exploratory Item Factor Analysis} \label{algorithm:1}}
\begin{enumerate}
    \item \textbf{Initialization} Input item responses $\mathbf{Y}$; dimension of latent space $P$; mini-batch size $M$; IW samples $R$; MC samples $S$; optimization hyperparameters $\eta$, $\beta_1$, and $\beta_2$; and starting values $\boldsymbol{\xi}_0 = (\boldsymbol{\theta}_0^{\top}, \boldsymbol{\psi}_0^{\top})^\top$
    
    \item At fitting iteration $t$, $t = 0, \ldots, T$:
    
    \begin{enumerate}
        \item \textbf{Computation} Randomly sample a mini-batch $\{\mathbf{y}_i\}_{i = 1}^M$; compute objective function value for respondent $i$, $i = 1, \ldots, M$:\par
        
        $\big( \boldsymbol{\mu}_i^\top, \log \boldsymbol{\sigma}_i^\top \big)^\top = \fnn_{\boldsymbol{\psi}_t}(\mathbf{y}_i)$ \par
        
        For IW sample $r$ and MC sample $s$, ${r = 1, \ldots, R}$, ${s = 1, \ldots, S}$:
        
            \qquad $\boldsymbol{\epsilon}_{i, r, s} \sim \mathcal{N} (\boldsymbol{\epsilon}_{i, r, s})$ \par
            
            \qquad $\mathbf{x}_{i, r, s} = \boldsymbol{\mu}_i + \boldsymbol{\sigma}_i \odot \boldsymbol{\epsilon}_{i, r, s}$ \par
            
            \qquad $\widetilde{\mathcal{L}}_1 = \sum_{j = 1}^J \sum_{k = 0}^{C_j - 1} \mathbbm{1}_k(y_{i,j}) \log \pi_{i,j,k} \;\;\;\enskip\quad \coloneqq \log p_{\boldsymbol{\theta}_t}(\mathbf{y}_i \mid \mathbf{x}_{i, r, s})$ \par
            
            \qquad $\widetilde{\mathcal{L}}_2 = \frac{1}{2} \sum_{p = 1}^P \big[\mu_{i, p}^2 + \sigma^2_{i, p} - 1 - \log \sigma^2_{i, p} \big] \coloneqq \dkl \big[\mathcal{N}(\mathbf{x}_{i, r, s} \mid \boldsymbol{\mu}_{i}, \boldsymbol{\sigma}_i^2 \mathbf{I}_P) \| \mathcal{N}( \mathbf{x}_{i, r, s} )\big]$ \par
            
            \qquad $w_{i, r, s} = \exp \big[ \widetilde{\mathcal{L}}_1 - \widetilde{\mathcal{L}}_2 \big]$
        
        $\iwelbo_i \approx \frac{1}{S} \sum_{s = 1}^S \Big[ \log \frac{1}{R} \sum_{r = 1}^R w_{i, r, s} \Big]$

        \item \textbf{Optimization} Update model parameters using AMSGrad: \par
        
        $\mathbf{g}_t = \frac{1}{M} \nabla_{\boldsymbol{\xi}_t} \sum_{i = 1}^M \iwelbo_i$ \par
        
        $\mathbf{m}_t = \beta_1 \mathbf{m}_{t - 1} + (1 - \beta_1) \mathbf{g}_t$ \par
        
        $\mathbf{v}_t = \beta_2 \mathbf{v}_{t - 1} + (1 - \beta_2) \mathbf{g}_t^2$ \par
        
        $\hat{\mathbf{v}}_t = \max(\hat{\mathbf{v}}_{t - 1}, \mathbf{v}_t)$ \par
        
        $\boldsymbol{\xi}_{t + 1} = \boldsymbol{\xi}_t - \eta \frac{\mathbf{m}_t}{\sqrt{\hat{\mathbf{v}}_t}}$ \par
    \end{enumerate}
    \item \textbf{Output} Return $\hat{\boldsymbol{\xi}} = \boldsymbol{\xi}_T$
\end{enumerate}
\end{algorithm}
\algcomment{IW = importance-weighted, MC = Monte Carlo.}

The inference model starting values $\boldsymbol{\psi}_0$ include a $P_l \times P_{l - 1}$ regression weight matrix $\mathbf{W}^{(l)}_0$ and a $P_l \times 1$ intercept vector $\mathbf{b}^{(l)}_0$ at FNN layers $l = 1, \ldots, L$. We initialize these parameters using a variant of Kaiming initialization (He et al., 2015), which has demonstrated good performance when applied to ANNs with asymmetric activation functions (e.g., the ELU function). Let $\mathcal{U} (a, b)$ denote a uniform density with lower bound $a$ and upper bound $b$. We randomly sample starting values as
\begin{equation} \label{eq:37}
        w_{p_1, p_{l - 1}, 0}^{(l)}, b_{p_l, 0}^{(l)} \sim \mathcal{U} \bigg( -\frac{1}{\sqrt{P_{l - 1}}}, \frac{1}{\sqrt{P_{l - 1}}} \bigg)
\end{equation}
for $p_l = 1, \ldots, P_l$, $p_{l - 1} = 1, \ldots, P_{l - 1}$, $l = 1, \ldots, L$. This initialization strategy often prevents the FNN hidden layer values from growing too large or too small at the start of fitting while accounting for the asymmetry of the ELU activation function around zero.

The starting values $\boldsymbol{\theta}_0$ include the $P \times 1$ factor loadings vector $\boldsymbol{\beta}_{j, 0}$ as well as the $(C_j - 1) \times 1$ intercept vector $\boldsymbol{\alpha}_{j, 0}$ for items $j = 1, \ldots, J$. We initialize the factor loadings using Xavier intialization (Glorot \& Bengio, 2010), which performs well when applied to ANNs with symmetric activation functions:
\begin{equation} \label{eq:38}
    \beta_{j, p, 0}^{(l)} \sim \mathcal{U} \bigg( -\sqrt{\frac{6}{J + P}}, \sqrt{\frac{6}{J + P}} \bigg),
\end{equation}
where $j = 1, \ldots, J$ and $p = 1, \ldots, P$. This approach stabilizes fitting in a manner similar to Kaiming initialization while accounting for the symmetry of the inverse logistic link function (i.e., equation~\ref{eq:8}) around zero. For $j = 1, \ldots, J$, we initialize the elements of $\boldsymbol{\alpha}_{j, 0}$ to an increasing sequence such that the cumulative density of logistic distribution between consecutive elements is the same (Christensen, 2019).

\subsection{Stabilizing Fitting and Checking Convergence}

We use a KL annealing strategy to avoid entrapment in local optima at the start of fitting (Bowman et al., 2016; Sønderby al., 2016). KL annealing multiplies the KL divergence term by $t / \tau$ for the first $\tau$ fitting iterations where $t = 0, \ldots, \tau - 1$. We conduct KL annealing for $\tau = \num{1000}$ fitting iterations for all models.

Once KL annealing is completed, we determine convergence similarly to Cremer et al. (2018). At each fitting iteration, we store the IW-ELBO computed for the associated mini-batch. After every $100$ fitting iterations, we compute the average of the previous $100$ mini-batch IW-ELBOs and compare this average to the previous best achieved average. If the best achieved average IW-ELBO does not improve after $100$ such comparisons, fitting is terminated.

It is sometimes necessary to assess whether different optimization runs have converged to equivalent stationary points. We conduct these checks using the estimated loadings matrices. We compare loadings matrices across runs by first rotating the factor solution using the Geomin oblique rotation method (Yates, 1988). Next, we invert factors if the sum of their loadings is negative (Asparouhov \& Muthén, 2009). We then select a reference matrix and find the column permutation of each comparison matrix that minimizes the element-wise mean squared error (MSE). Finally, we compute Tucker's congruence coefficient between the permuted matrices (Lorenzo-Seva \& ten Berge, 2006). Solutions with congruence coefficients larger than $0.98$ are deemed equivalent (MacCallum et al., 1999). We note that to compare factor correlation solutions, the same inversion and permutation procedure is applied to both columns and rows of the estimated factor correlation matrices.

\subsection{Tuning Hyperparameters}

Inference model hyperparameters include the number and size of the FNN hidden layers. After some experimentation, we found that performance was relatively insensitive to these values. We therefore use a single hidden layer for all models and set the hidden layer size to a value close to the mean of the input layer size and twice the latent dimension $P$. This choice is based on the observation that ``the optimal size of the hidden layer is usually between the size of the input and size of the output layers'' (Heaton, 2008).

Optimization hyperparameters include the forgetting factors for the gradient and squared gradient, $\beta_1$ and $\beta_2$; the learning rate $\eta$; and the mini-batch size $M$. We set $\beta_1 = 0.9$ and $\beta_2 = 0.999$, which are default values typically recommended in practice (Reddi et al., 2018). We set $\eta = 0.005$ for most models. For some models with many factors and many items, the IW-ELBO diverged, so we set $\eta = 0.0025$. This approach is based on the observations that $\eta \leq 0.005$ typically performs well for adaptive SG methods and $\eta$ should be reduced if the objective function diverges (Bengio, 2012). Keskar et al. (2017) note that mini-batch sizes $M \geq 32$ perform well in many applications, and Bengio (2012) notes that $M$ mostly impacts time to convergence rather than model performance. We therefore set $M = 32$ as a default value for all analyses.\footnote{$M$ is typically set to a power of $2$ to reduce fitting times by facilitating GPU (or CPU) memory allocation (Goodfellow et al., 2016).}

Setting the number of IW samples $R$ typically does not require extensive tuning but does require some consideration. Empirically, we found that increasing $R$ increases the accuracy of $\boldsymbol{\theta}$ estimates. However, even for small $R$ (e.g., $R = 5$), amortized IWVI typically yields comparable $\boldsymbol{\theta}$ estimates to state-of-the-art MML estimation procedures in less time. We also found that computational efficiency increases for small $R$ then decreases as $R$ grows large (e.g., $R = 25$). These results suggest that $R$ may be chosen according to whether (1) computational efficiency or (2) highly accurate $\boldsymbol{\theta}$ estimates are desired: If (1), choose some small $R$ for which the algorithm converges quickly; if (2), choose the largest $R$ for which the algorithm converges in a reasonable amount of time.

The main hyperparameter that requires tuning is the latent dimension $P$. We tried tuning $P$ using a pseudo-likelihood Bayesian information criterion (pseudo-BIC; Erosheva et al., 2007) as well as using a more computationally intensive $5$-fold cross-validation (CV) approach (details available upon request) but found that both approaches performed poorly as $N$ increased. We therefore use a more subjective scree plot approach based on the Monte Carlo CV method described by Hui et al. (2017). To construct each scree plot, we first create a holdout set by randomly sampling some percentage of the item responses without replacement. Let $\Omega$ denote the index set for the item responses in the holdout set and let $\Omega'$ denote the indices of the item responses excluding the holdout set. For a fixed $P$, we fit the model using the item responses indexed by $\Omega'$. We denote the fitted parameters so obtained as $\hat{\boldsymbol{\xi}} = (\hat{\boldsymbol{\theta}}^\top, \hat{\boldsymbol{\psi}}^\top)^\top$. Treating the IW-ELBO with $R = 5000$ IW samples as a close approximation to the true log-likelihood (Cremer et al., 2018), we predict the approximate log-likelihood for the holdout set as
\begin{equation} \label{eq:39}
    \tilde{\ell}(P) = \sum_{i \in \Omega} \bigg[ \log \frac{1}{5000} \sum_{r = 1}^{5000} \frac{p_{\hat{\boldsymbol{\theta}}}(\mathbf{x}_{i, r}, \mathbf{y}_i)}{q_{\hat{\boldsymbol{\psi}}}(\mathbf{x}_{i, r} \mid \mathbf{y}_i)} \bigg],
\end{equation}
which corresponds to equation~\ref{eq:35} with $S = 1$ and $\boldsymbol{\xi} = \hat{\boldsymbol{\xi}}$. After performing the above procedure for several successive values of $P$, the scree plot is constructed by plotting $-\tilde{\ell}(P)$ against increasing $P$. The latent dimension coinciding with an ``elbow'' in the plot may be selected. We note that this approach differs from traditional scree plots in that we plot predicted approximate log-likelihoods rather than eigenvalues, although both approaches are interpreted similarly (i.e., look for the ``elbow''). We empirically evaluate this approach in the following section.

\section{Numerical Examples} \label{section:6}

Models were programmed with the machine learning library PyTorch (Version 1.1.6; Paszke et al., 2019) and were fitted on a laptop computer with a 2.8 GHz Intel Core i7 CPU and 16 GB of RAM. Although GPU computing is directly supported in PyTorch and often speeds up fitting, we opted for CPU computing to enable fairer comparisons with other methods and to assess performance using hardware more typically available for psychology and education research. All code is available as online supplemental material.

\subsection{Application to a Big-Five Personality Questionnaire} \label{section:6.1}

We first demonstrate amortized IWVI via an empirical example intended to: (1) showcase IWVI's computational efficiency when analyzing large-scale item response data and (2) obtain reasonable population values for conducting simulation studies. Specifically, we analyze $\num{1015342}$ responses to Goldberg's (1992) 50 Big-Five Factor Marker (FFM) items from the International Personality Item Pool (IPIP; Goldberg et al., 2006) downloaded from the Open-Source Psychometrics Project (\url{https://openpsychometrics.org/}). The IPIP-FFM items were designed to assess respondents' levels of five personality factors: \textit{Conscientiousness}, \textit{openness}, \textit{emotional stability}, \textit{agreeableness}, and \textit{extraversion}. Empirical Big-Five studies often yield substantial factor inter-correlations (e.g., Biesanz \& West, 2004), so we permitted correlated factors by applying the Geomin oblique rotation method to all fitted loadings matrices. Each of the five factors included $10$ five-category items anchored by ``Disagree'' (1), ``Neutral'' (3), and ``Agree'' (5). Item responses were recoded as necessary so that the highest numerical value of the response scale indicated a high level of the corresponding factor. After pre-processing the data (details available upon request), our final sample size was $N = \num{515708}$ responses.

\begin{figure}
	\centering
	\includegraphics[width=0.55\textwidth]{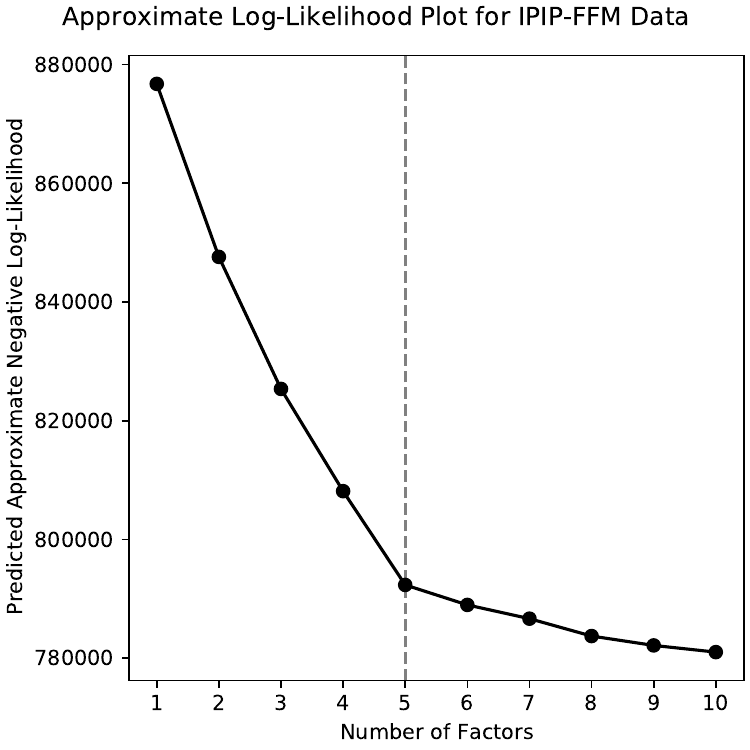}
	\caption{Scree plot of predicted approximate negative log-likelihood as a function of the number of latent factors. The ``elbow'' at $5$ factors is marked with a dotted line.} \label{fig:3}
\end{figure}

Computation was carried out following the procedures described in Section~\ref{section:5}. A scree plot of $-\tilde{\ell}(P)$ computed on a holdout set of $2.5\%$ of observations for $P \in \{ 1, \ldots, 10\}$ (Figure~\ref{fig:3}) demonstrated an ``elbow'' at $P = 5$, suggesting that $5$ latent factors accounted for most of the correlation between item responses. We set the inference model hidden layer size to $130$ (i.e., the mean of the input layer size and $2P$) and the learning rate to $\eta = 0.005$. We set the number of IW samples $R = 5$ to demonstrate the importance-weighting approach. We fitted the full data set $100$ times to assess the replicability of the parameter estimates across random seeds. Only equivalent factor solutions were compared.

\begin{figure}
	\centering
	\includegraphics[width=0.9\textwidth]{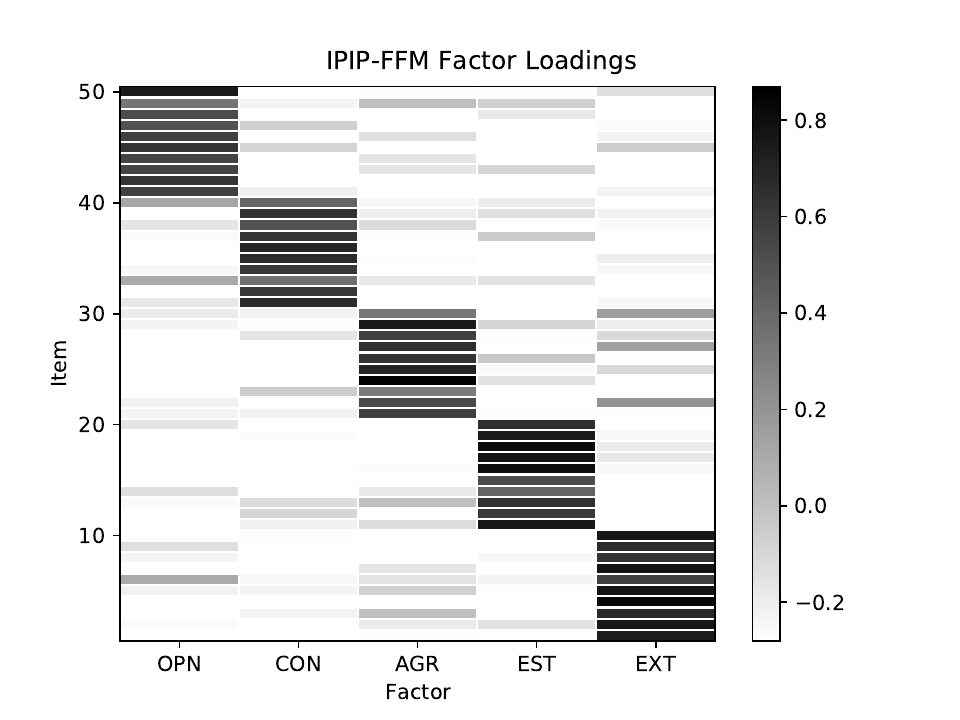}
	\caption{Heat map of factor loadings for IPIP-FFM items. EXT = extraversion, EST = emotional stability, AGR = agreeableness, CON = conscientiousness, OPN = openness.} \label{fig:4}
\end{figure}

We report results from the fitted model that attained the highest IW-ELBO. Figure~\ref{fig:4} contains a heatmap of the Geomin-rotated factor loadings estimates, which fit with the expected five factor structure. Factor correlations in Table~\ref{table:1} also fit with the typical finding that emotional stability is negatively correlated with the other factors. Notably, fitting was fast: Mean fitting time across random seeds was $170$ seconds ($\mathrm{\mathit{SD}} = 47$ seconds). Further, parameter estimates were fairly stable: Across random seeds, mean loadings root-mean-square error (RMSE) was $0.018$ ($\mathrm{\mathit{SD}} = 0.006$), mean intercepts RMSE was $0.042$ ($\mathrm{\mathit{SD}} = 0.018$), and mean factor correlation RMSE was $0.028$ ($\mathrm{\mathit{SD}} = 0.009$).

\begin{table}
\centering
\addtolength{\tabcolsep}{3pt} 
\caption{Factor correlations for IPIP-FFM data set.} \label{table:1}

    \begin{tabular}{lS[table-format = .2,add-integer-zero=false,detect-weight]S[table-format = .2,add-integer-zero=false,detect-weight]S[table-format = .2,add-integer-zero=false,detect-weight]S[table-format = .2,add-integer-zero=false,detect-weight]S[table-format = .2,add-integer-zero=false,detect-weight]}
        \toprule
             & \multicolumn{5}{c}{Factor} \\ \cmidrule{2-6}
        Factor & {$1$} & {$2$} & {$3$} & {$4$} & {$5$} \\
        \midrule
        1. Extraversion & 1.00 & & & & \\
        2. Emotional Stability & -.20 & 1.00 & & & \\
        3. Agreeableness & .15 & -.01 & 1.00 & & \\
        4. Conscientiousness & .08 & -.22 & .11  & 1.00 & \\
        5. Openness & .12 & -.05 & .10 & -.01 & 1.00 \\
        \midrule[\heavyrulewidth]
    \end{tabular}
\end{table}

\subsection{Simulation Studies}

\subsubsection{Evaluation of the Importance-Weighting Procedure}

In this study, we investigate amortized IWVI's performance in terms of parameter recovery and computational efficiency as the number of IW samples $R$ increases (i.e., as the approximation to the marginal likelihood improves). We consider $R = 1$, $5$, and $25$. The first setting uses the ELBO objective, while the latter settings use the IW-ELBO objective. Data generating loadings, intercepts, and factor correlations are rounded estimates from the IPIP-FFM example in section~\ref{section:6.1}. We set $P = 5$, $J = 50$, and $C_j = 5$ for $j = 1, \ldots, J$. Each factor loads on ten items with cross loadings set to zero to produce a perfect simple structure. To investigate IWVI's performance as the sample size increases, we conduct $100$ replications of simulation for $N = \num{500}$, $\num{1000}$, $\num{2000}$, and $\num{10000}$. This leads to $12$ different simulation settings for all possible combinations of $R$ and $N$. We also assessed the model selection performance of the scree plot approach by plotting $-\tilde{\ell} (P)$ computed on a holdout set of $20\%$ of observations for $P = 2, \ldots, 8$ at each replication. All inference model and optimization hyperparameters from section~\ref{section:6.1} were reused for these analyses.

\begin{figure}
	\centering
	\includegraphics[width=\textwidth]{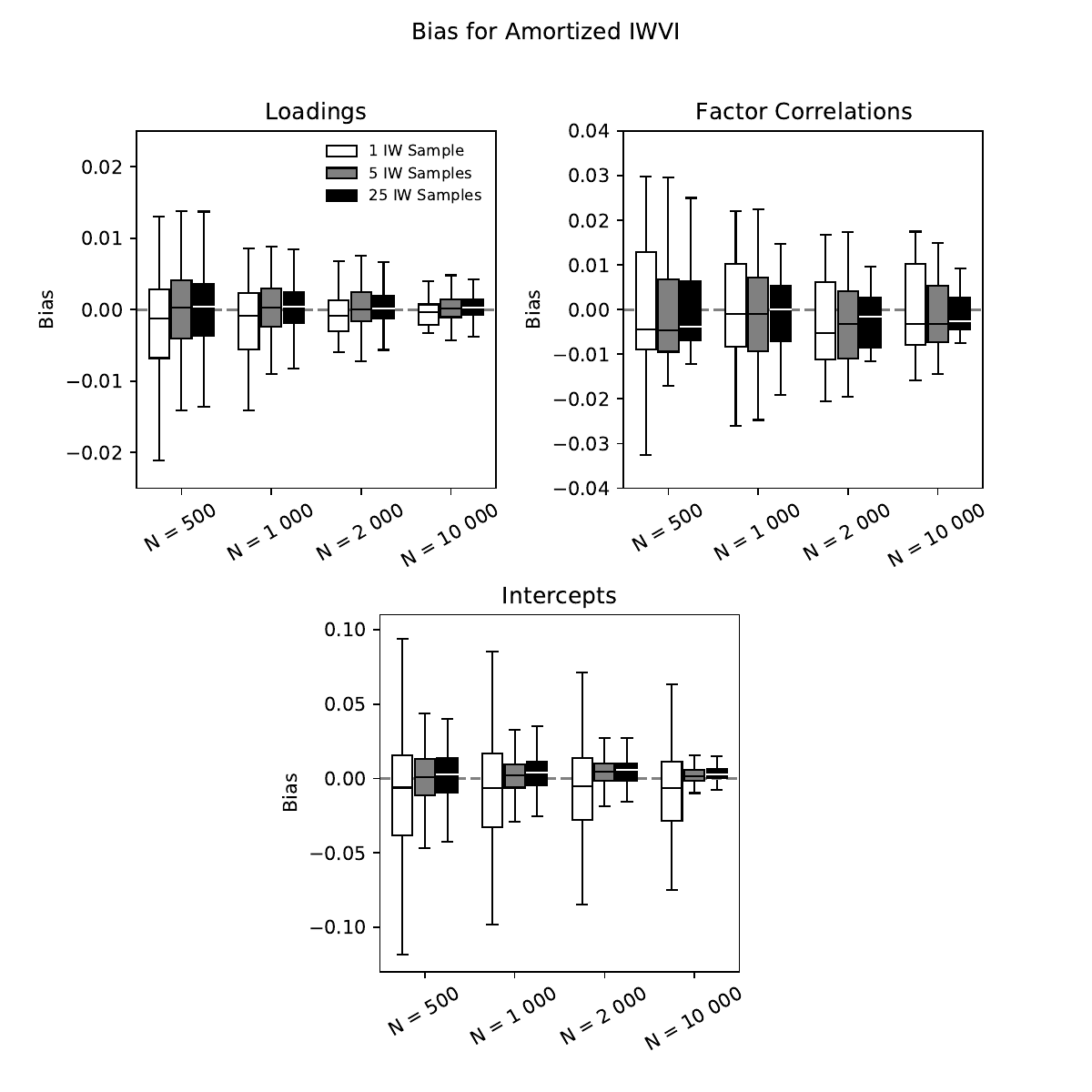}
	\caption{Parameter bias for amortized importance-weighted variational inference (IWVI) computed based on $100$ replications of simulation. Three settings for the number of importance-weighted (IW) samples are compared.} \label{fig:5}
\end{figure}

\begin{figure}
	\centering
	\includegraphics[width=\textwidth]{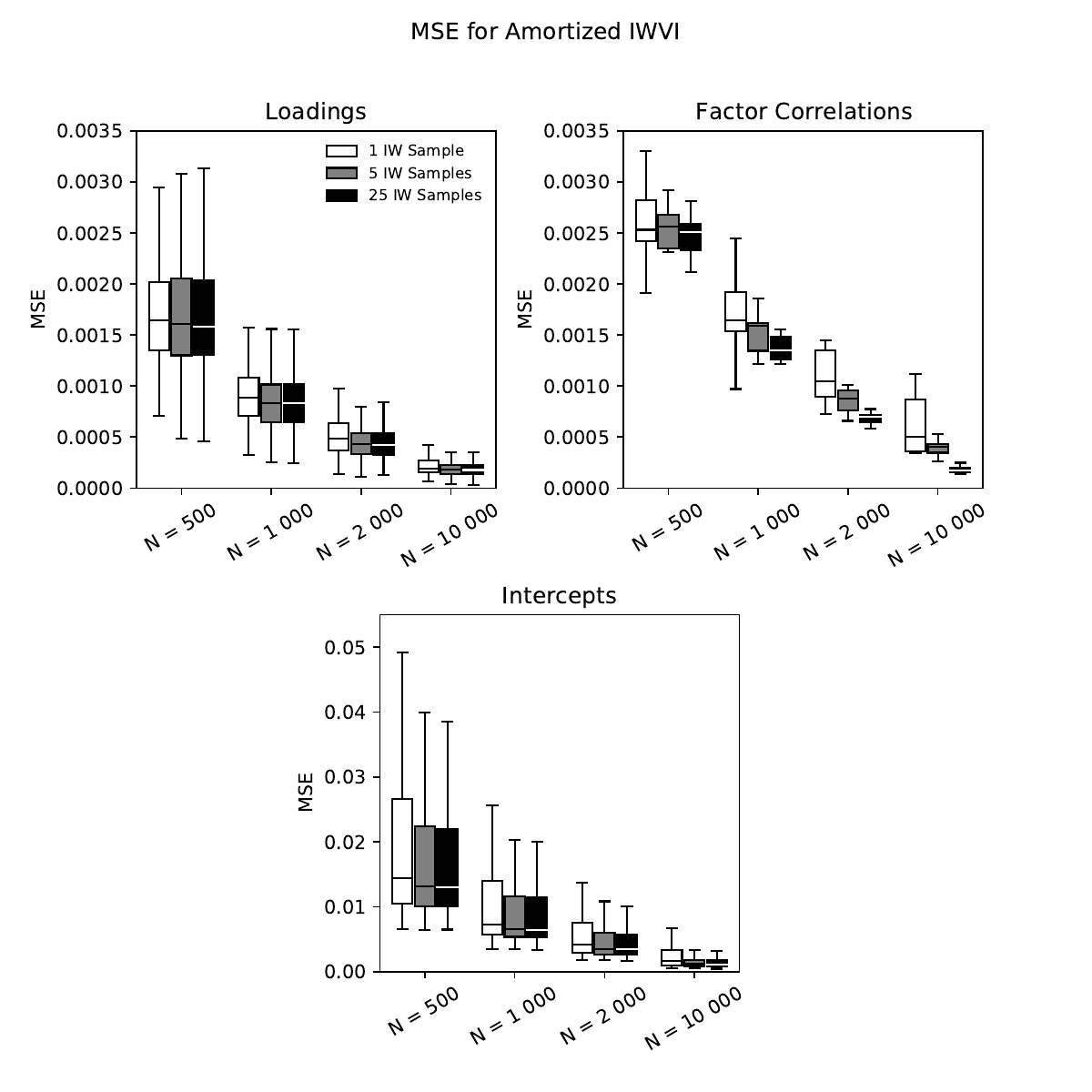}
	\caption{Parameter mean squared error (MSE) for amortized IWVI computed based on $100$ replications of simulation.} \label{fig:6}
\end{figure}

To assess parameter recovery, we computed bias for each parameter as the mean deviation of the estimated parameter from the data generating parameter across replications:
\begin{equation} \label{eq:40}
    \bias(\hat{\xi}, \xi) = \frac{1}{100} \sum_{a = 1}^{100} [ \hat{\xi}^{(a)} - \xi ],
\end{equation}
where $\hat{\xi}^{(a)}$ is the estimated parameter at replication $a$ and $\xi$ is the data generating parameter. Figure~\ref{fig:5} uses boxplots to summarize the parameter biases separately for the factor loadings, factor correlations, and intercepts. All estimates become more accurate as $N$ increases. Intercepts and factor correlation estimates become more accurate with increasing $R$ but exhibit slight bias across $N$ settings. Loadings estimates are nearly unbiased and either become more accurate or obtain comparable accuracy as $R$ increases. We also computed MSE for each parameter (i.e., by squaring the summands in equation~\ref{eq:40}). Results are summarized using boxplots in Figure~\ref{fig:6}. For each $R$ setting, parameter MSE quickly decreases toward zero with increasing $N$. Increasing $R$ tends to decrease MSE for each $N$ setting, with factor correlation estimates demonstrating particularly large improvements as $R$ increases.

\begin{figure}
	\centering
	\includegraphics[width=0.55\textwidth]{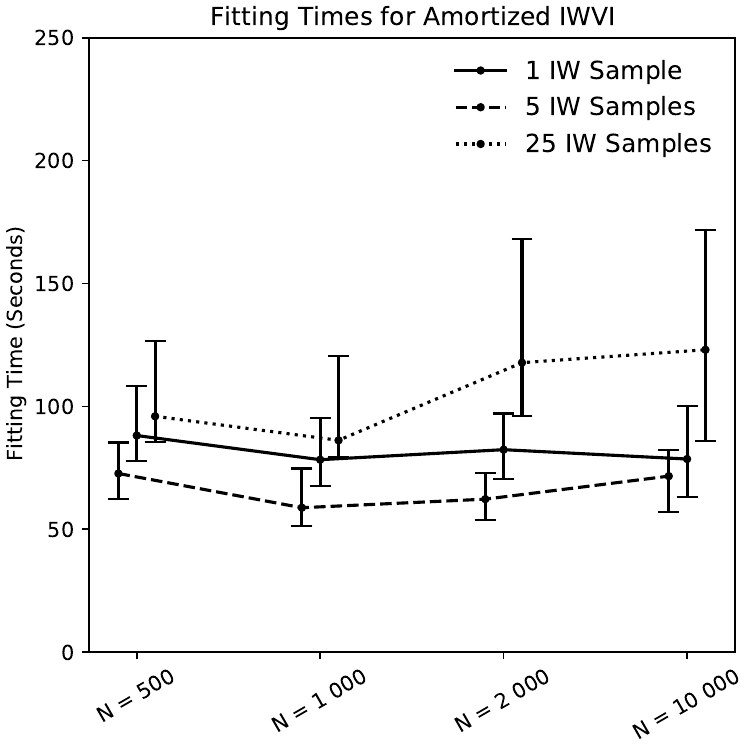}
	\caption{Fitting times for amortized IWVI across $100$ replications of simulation. For all line plots in this work, points indicate medians while error bars indicate $25\%$ and $75\%$ quantiles.} \label{fig:7}
\end{figure}

Figure~\ref{fig:7} contains line plots of fitting times for each simulation setting across replications. Increasing $R$ from $1$ to $5$ leads to a decrease in median fitting time (around $80$ seconds to around $65$ seconds), while increasing $R$ from $5$ to $25$ leads to a moderate increase ($R = 25$ takes around $90$ to $120$ seconds). Median fitting times for $R = 1$ and $5$ were essentially constant as $N$ increased. The median fitting time for $R = 25$ increased around $20$ seconds from $N = \num{1000}$ to $\num{2000}$, although absolute fitting times for this $R$ setting were never unreasonably large. These results highlight the scalability of AMSGrad to very large data sets.

\begin{figure}
	\centering
	\includegraphics[width=0.55\textwidth]{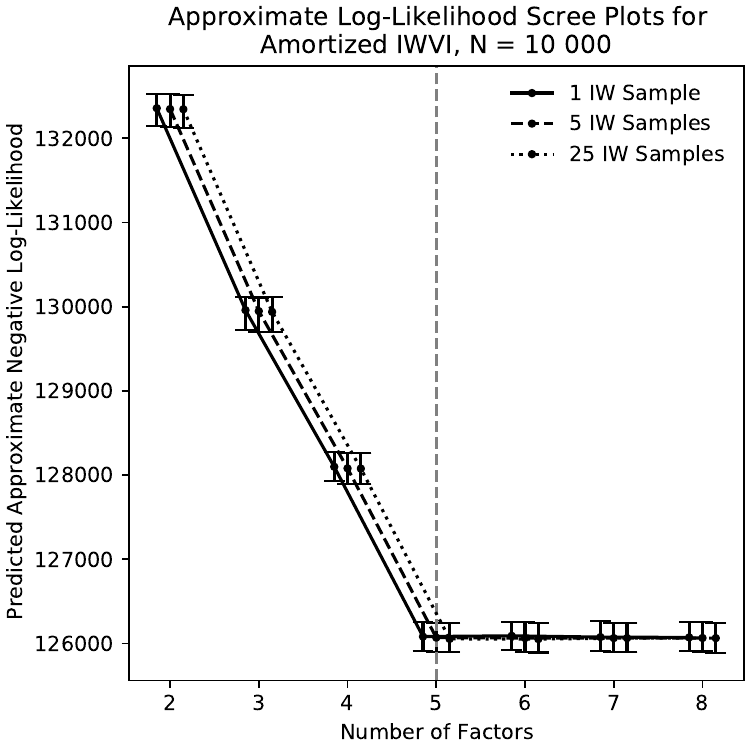}
	\caption{Approximate log-likelihood scree plots for amortized IWVI constructed for simulation settings with~${N = \num{10000}}$. The ``elbows'' at $5$ factors are marked with a horizontal dotted line.} \label{fig:8}
\end{figure}

To assess factor score estimation accuracy at each replication, we first obtained expected \textit{a posteriori} (EAP) factor score estimates for all models. For $R = 1$, we obtained EAPs by computing the approximate LV posterior mean $\boldsymbol{\mu}$ for each respondent. For $R = 5$ and $25$, we obtained EAPs for each respondent by averaging $S = 50$ Monte Carlo samples drawn from $q_{\boldsymbol{\psi}}^{\text{IW}}(\mathbf{x} \mid \mathbf{y})$ using sampling-importance-resampling (for details, see Cremer et al., 2017). After rotating the scores and applying the inversion and column permutation procedure used to compare loadings solutions, we computed the correlation between the true and estimated scores for each latent factor. Estimates were accurate: For fixed $R$, correlations ranged from $0.88$ to $0.95$ and tended to increase with increasing $N$. Correlations also increased slightly with increasing $R$ and fixed $N$. The scree plot approach to tuning the latent dimension $P$ appeared to perform well across simulation settings. Figure~\ref{fig:8} presents scree plots for simulation settings where $N = \num{10000}$, which possess sharp ``elbows'' at $P = 5$. Median $-\tilde{\ell}(P)$ values decreased slightly with increasing $R$, indicating that importance sampling helped models obtain slightly better fit to previously unseen data.  Plots for other $N$ settings were nearly identical and are not shown.

\subsubsection{Comparison to MH-RM} \label{section:6.2.2}

In this study, we compare amortized IWVI to the MML estimator implemented via the MH-RM algorithm. We note that the stEM algorithm is somewhat faster than MH-RM (Zhang, Chen, \& Liu, 2020) and may therefore be a suitable alternative comparison method. However, given that MH-RM is relatively widely used and that stEM has only been implemented for the M2PL, we choose MH-RM for these analyses. MH-RM is implemented via the R package mirt (Version 1.32.1; Chalmers, 2012). Comparing the computational efficiency of the proposed approach and MH-RM is therefore fair in the sense that both mirt and Pytorch core functions are written in C++ and comparisons are conducted on the same computer.

We compare these methods in the high-dimensional setting where $P = 10$, $J = 100$, and $C_j = 5$ for $j = 1, \ldots, J$. Data generating parameters are again rounded estimates from the IPIP-FFM example. We set the parameters for items $51$-$100$ equal to the parameters for items $1$-$50$. We construct the factor correlation matrix as a $10 \times 10$ block diagonal matrix with main-diagonal blocks equal to the rounded IPIP-FFM estimates and zeros elsewhere. Results of the previous simulation suggest that choosing a small $R > 1$ increases both estimation accuracy and computational efficiency, so we set $R = 5$ for these analyses. All other hyperparameters are set as in previous sections. MH-RM hyperparameters were set to the mirt package defaults. We conduct $100$ replications of simulation for $N = \num{1000}$, $\num{2000}$, $\num{5000}$, and $\num{10000}$.

\begin{figure}
	\centering
	\includegraphics[width=\textwidth]{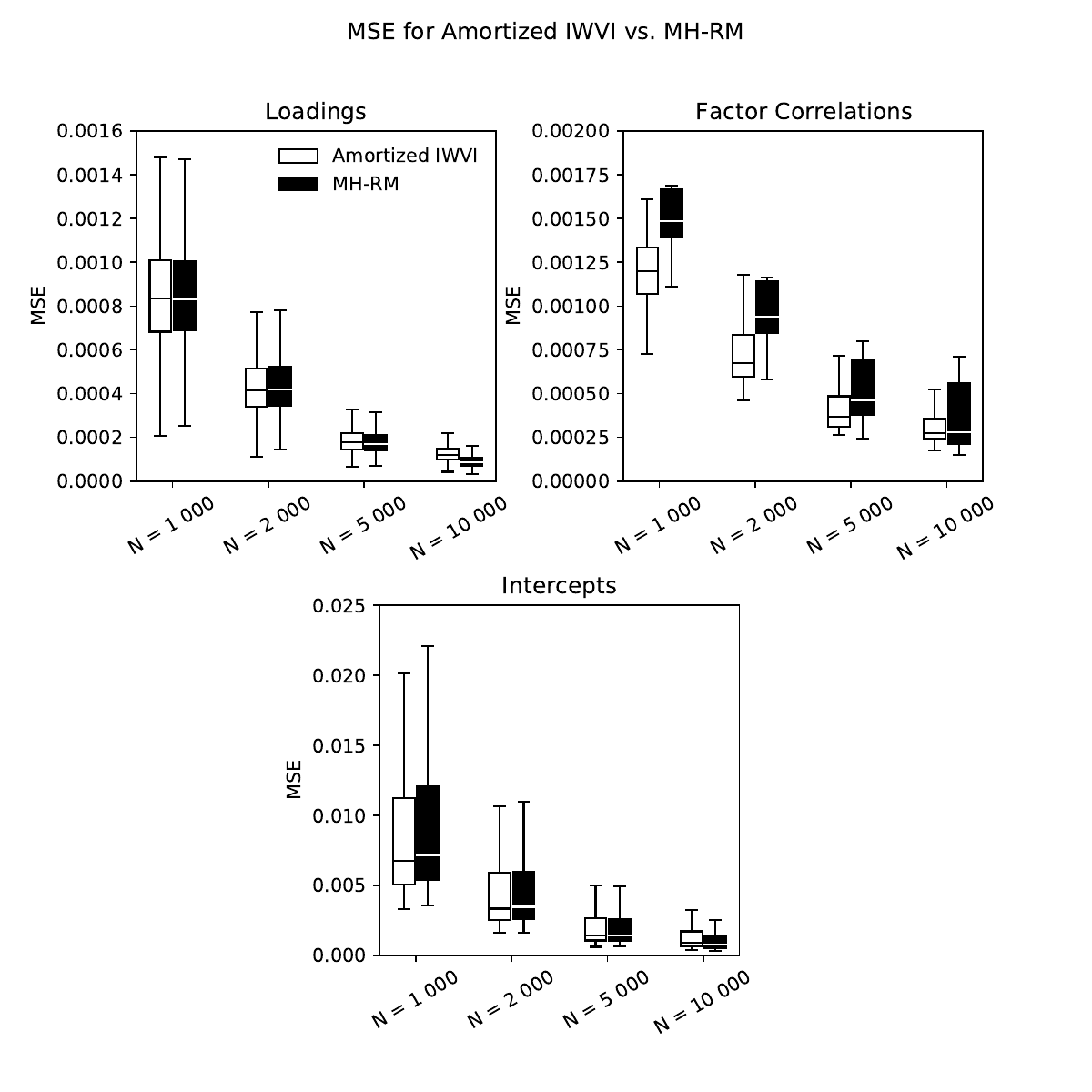}
	\caption{MSE for amortized importance-weighted variational inference (IWVI) and the marginal maximum likelihood estimator computed based on $100$ replications of simulation. MH-RM = Metropolis-Hastings Robbins-Monro.} \label{fig:9}
\end{figure}

\begin{figure}
	\centering
	\includegraphics[width=0.6\textwidth]{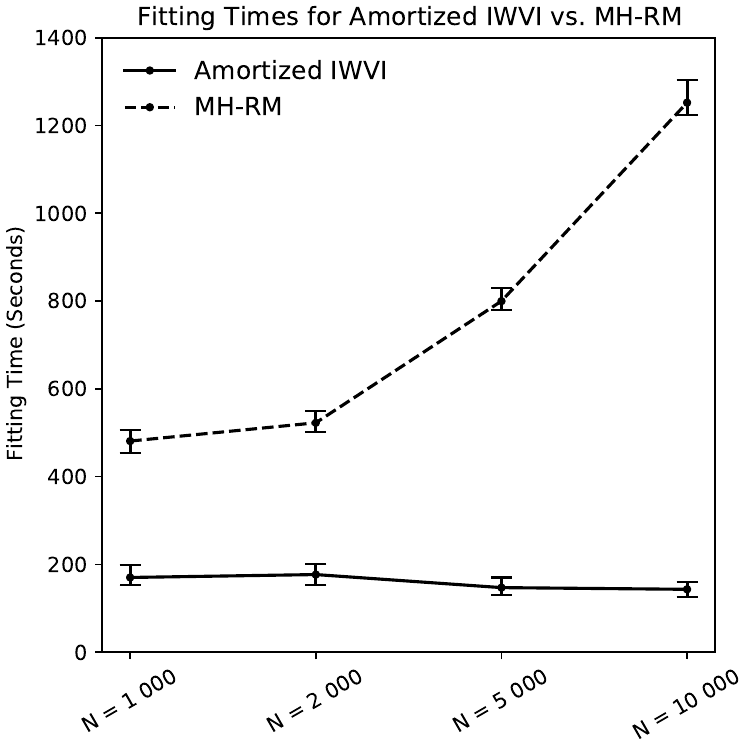}
	\caption{Fitting times for amortized IWVI and the marginal maximum likelihood estimator computed based on $100$ replications of simulation.} \label{fig:10}
\end{figure}

Results are shown in Figures~\ref{fig:9} and~\ref{fig:10}. Both methods obtain comparable loadings and intercepts estimates for $N \leq \num{5000}$. MH-RM's loadings and intercepts estimates are slightly more accurate when $N = \num{10000}$, although both methods are very accurate in this setting. Amortized IWVI produces more accurate factor correlation estimates across $N$ settings. Additionally, IWVI is much faster than MH-RM: MH-RM's median fitting time increases from $8$ minutes for $N = \num{1000}$ to $21$ minutes for $N = \num{10000}$, whereas IWVI's median fitting time stays around $3$ minutes regardless of $N$.

\subsubsection{Comparison to CJMLE}

We also compare amortized IWVI to CJMLE in the setting where $N$ and $J$ increase simultaneously. Chen, Li, and Zhang (2019) empirically show that the MML estimator implemented via MH-RM performs poorly when both $N$ and $J$ increase and that CJMLE attains much faster convergence via an alternating minimization algorithm. CJMLE is implemented in the R package mirtjml (Version 1.4; Zhang, Chen, \& Li, 2019) and has core functions written in C++. Although CJMLE computation may be parallelized, we compare methods using a single core to ensure fairness.

We again set $P = 10$ and consider $(N, J) = (\num{2000}, 100)$, $(\num{10000}, 200)$, $(\num{50000}, 300)$, and $(\num{100000}, 400)$. CJMLE is only implemented for the M2PL, so we set $C_j = 2$ for $j = 1, \ldots, J$. Data generating item parameters are again set by repeating the IPIP-FFM item parameters. For example, when $J = 400$, we repeat the parameters for items $1$-$50$ seven times to get the parameters for items $51$-$400$. Since each item needs only a single intercept, we randomly select an intercept from the four fitted IPIP-FFM intercepts for each item. The factor correlation matrix from section~\ref{section:6.2.2} is reused. We set $\eta = 0.005$ for settings with $J \leq \num{200}$ but set $\eta = 0.0025$ when $J \geq \num{300}$ because these larger models sometimes diverged otherwise. Other hyperparameters are set similarly to those in section~\ref{section:6.2.2}.

\begin{figure}
	\centering
	\includegraphics[width=\textwidth]{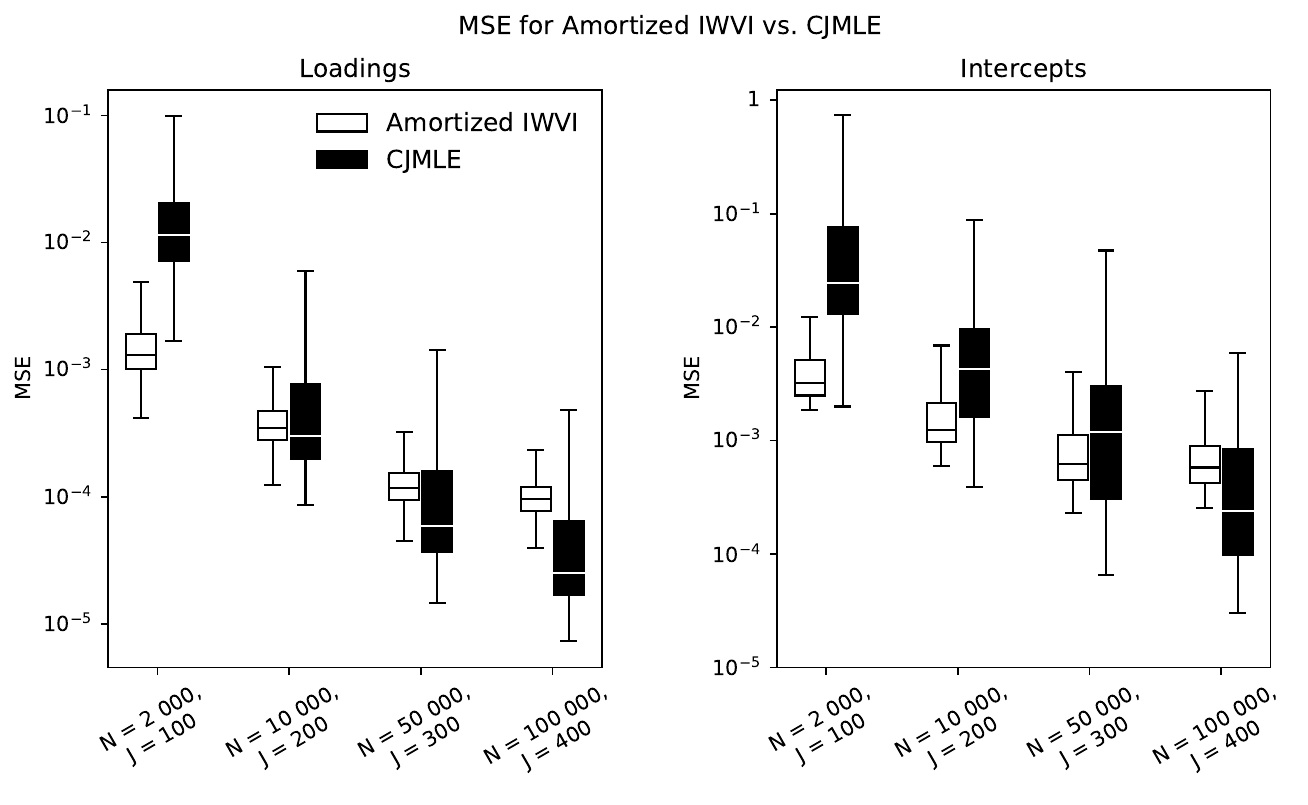}
	\caption{MSE for amortized IWVI and the constrained joint maximum likelihood estimator (CJMLE) computed based on $100$ replications of simulation.} \label{fig:11}
\end{figure}

\begin{figure}
	\centering
	\includegraphics[width=0.6\textwidth]{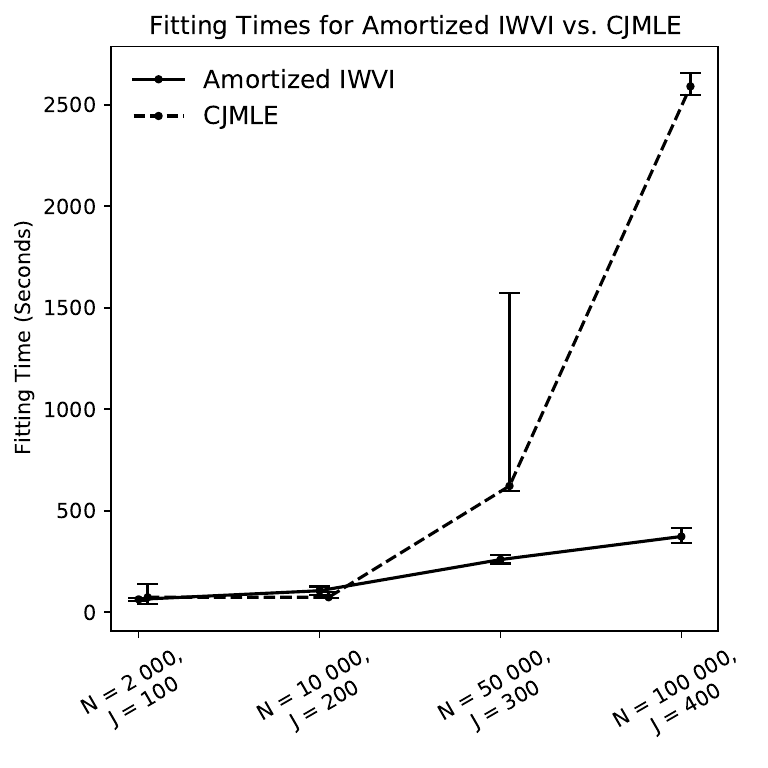}
	\caption{Fitting times for amortized IWVI and CJMLE computed based on $100$ replications of simulation.} \label{fig:12}
\end{figure}

Results are presented in Figures~\ref{fig:11} and~\ref{fig:12}. Unlike the MML estimator, amortized IWVI performs better as $N$ and $J$ increase and attains comparable accuracy to CJMLE. CJMLE estimates loadings and intercepts inaccurately in the smallest $(N, J)$ setting but is more accurate than IWVI in the highest setting (although both methods are accurate when $N \geq \num{50000}$). Factor correlation estimates are not reported because CJMLE treats the LVs as fixed effects. Importantly, IWVI is always comparably fast or faster than CJMLE. IWVI's median fitting time is $63$ seconds when $(N, J) = (\num{2000}, 100)$ and increases to just over $6$ minutes when $(N, J) = (\num{100000}, 400)$, whereas CJMLE's median fitting time increases from $73$ seconds to over $43$ minutes in the same settings. When $(N, J) = (\num{50000}, 300)$, CJMLE sometimes took around $\num{1500}$ seconds to converge rather than around $\num{600}$ seconds (i.e., the median fitting time). This is possibly due to CJMLE converging to different local optima of the joint likelihood function. We note that CJMLE may achieve a significant speedup using parallel computing, although IWVI may achieve a similar speedup using a GPU.

\section{Extensions} \label{section:7}

We now briefly consider a variety of interesting ways in which application of amortized IWVI could be expanded.

\subsection{Confirmatory Item Factor Analysis}

Confirmatory IFA is useful when sufficient prior theory exists to posit a specific factor structure for the items. More precisely, the measurement design may be indicated by a pre-specified $J \times P$ matrix $\mathbf{Q}$ with entries $q_{j, p} \in \{0, 1\}$ such that $q_{j, p} = 1$ if item $j$ measures factor $p$ (i.e., $\beta_{j, p}$ is freely estimated) and $q_{j, p} = 0$ otherwise (i.e., $\beta_{j, p}$ is set to zero). Anderson and Rubin (1957) provide sufficient conditions on $\mathbf{Q}$ to ensure the model is identified. The deep learning algorithm discussed here may be used to conduct confirmatory IFA by ensuring that factor loadings are either freely estimated or set to zero as specified in $\mathbf{Q}$ (Curi et al., 2019). In the confirmatory setting, it is also sometimes of interest to impose constraints on the factor covariance matrix $\boldsymbol{\Sigma}$. Letting $\boldsymbol{\Sigma} = \mathbf{L} \mathbf{L}^\top$ where $\mathbf{L}$ is a lower triangular matrix, we can estimate $\mathbf{L}$ using a hyperspherical parameterization (Pinheiro \& Bates, 1996; Rapisarda et al., 2007), which enables unconstrained estimation of a variety of structured correlation matrices and has similar computational efficiency to estimating $\mathbf{L}$ directly (e.g., Ghosh et al., 2020; Tsay \& Pourahmadi, 2017).

\subsection{Regularized Exploratory Item Factor Analysis}

Regularization has been proposed as a viable alternative to factor rotation for both exploratory linear factor analysis (e.g., Choi et al., 2010; Hirose \& Konishi, 2012; Hirose \& Yamamoto, 2014) and exploratory IFA (Hui et al., 2018; Sun et al., 2016). Many regularization approaches automatically rotate the factors to produce a sparse loadings structure. The regularized, amortized, importance-weighted variational estimator is obtained by solving the optimization problem
\begin{equation} \label{eq:41}
   \boldsymbol{\xi}^* = \argmax_{\boldsymbol{\xi}} \bigg[ \sum_{i = 1}^N \iwelbo_{i} - \mathcal{P}(\mathbf{B}) \bigg],
\end{equation}
where $\mathbf{B}$ is a $J \times P$ factor loadings matrix whose $j^\mathrm{th}$ row is $\boldsymbol{\beta}_j$ and $\mathcal{P}$ is a penalty function that is potentially non-smooth and non-convex. This optimization problem may be solved using a proximal version of AMSGrad based on the ProxGen procedure developed by Yun et al. (2020), which is guaranteed to converge to a local stationary point when mild conditions are satisfied.

\subsection{Flexible Latent Density Estimation}

Recent work by Monroe (2014) aims to relax the assumption that the LVs are multivariate normally distributed. An alternative approach developed recently in the deep learning literature is based on the concept of normalizing flows (NFs; Tabak \& Turner, 2012; Tabak \& Vanden-Eijnden, 2010). NFs apply a sequence of invertible mappings parameterized by ANNs that aims to transform a simple base density into an arbitrarily complicated density. Since the mappings are invertible, the transformed density can be explicitly evaluated via the change of variables formula. NFs scale well to high-dimensional spaces and may be used to increase the flexibility of AVI by building complicated latent prior or posterior distributions (e.g., Huang et al., 2018; Kingma et al., 2016; Rezende \& Mohamed, 2015).

\subsection{Nonlinear Factor Analysis}

The full IWAE may be viewed as a model for nonlinear factor analysis (Yalcin \& Amemiya, 2001) of the form
\begin{equation} \label{eq:42}
  \mathbf{y}_i  = g(\mathbf{x}_i) + \boldsymbol{\varepsilon}_i,
\end{equation}
for $i = 1, \ldots, N$ where $g$ is an arbitrary nonlinear function and $\boldsymbol{\varepsilon}_i$ is the $i^\mathrm{th}$ $J \times 1$ vector of errors. In the IWAE, $g$ is approximated using an ANN. This corresponds to approximating the inverse link function between observed and latent variables while keeping the latent density fixed (Wu et al., 2020). This approach is typically less interpretable than approximating the latent density and fixing $g$, which provides equivalent model fit (e.g., Woods \& Thissen, 2006).

\section{Discussion} \label{section:8}

In this paper, we explored the suitability of an amortized importance-weighted variational inference algorithm for exploratory IFA. Numerical studies highlighted several benefits of the proposed approach. Analysis of a large-scale Big-Five personality factors data set yielded fast results that aligned with existing psychological theory across random starts. Our simulations suggested that, unlike other estimators, the amortized importance-weighted variational estimator performs comparably to state-of-the-art estimators in both the setting where the number of observations increases and in the setting where the number of items and the number of observations simultaneously increase. Amortized IWVI converges faster than existing approaches when optimized using the adaptive stochastic gradient algorithm AMSGrad, particularly with large-scale data. Factor score estimates were accurate and improved with increasing sample size. The sampling-based initialization procedures appeared to mitigate problems associated with convergence to local optima and performed comparably to the more computationally intensive initialization procedures used by MH-RM and CJMLE.

Two practical considerations not discussed here are standard errors (SEs) and missing data. Hui et al. (2017) note that for the former, approximate SEs may be obtained by evaluating the observed information matrix at the estimates $\hat{\boldsymbol{\theta}}$ obtained by maximizing the IW-ELBO. Since this matrix has a block diagonal structure, it may be block-wise inverted to produce the covariance matrix from which SEs can be calculated. We note that SEs will likely be quite small for the large-scale applications considered here. Mattei and Frellsen (2019) discuss a simple approach to handling missing-at-random data in amortized IWVI that can be easily applied to the models considered here.

The proposed approach has several limitations. The main practical difficulty we encountered was tuning the number of latent factors $P$. Although we tried tuning $P$ using objective methods such as computing a pseudo-BIC and conducting $5$-fold CV, these methods typically failed for large sample sizes, possibly due to log-likelihood approximation error exceeding sampling error. We therefore used subjective log-likelihood scree plots to tune $P$. Research is needed to develop objective criteria for selecting the latent dimension in large samples. In the meantime, the more subjective scree plot approach used here as well as approaches such as parallel analysis and retaining theoretically meaningful factors may serve as practical substitutes.

As noted by Hui et al. (2017), substantial theoretical work remains to be done to show that variational approximations produce consistent, asymptotically normal estimators and to obtain their rates of convergence. The importance sampling approach explored here provides a theoretical link between VI and MML estimation. Our simulations showed that obtaining a better approximation to the true marginal likelihood using importance sampling increases parameter estimation accuracy, although intercepts and factor correlation estimates exhibited some bias. Future theoretical work may support our empirical results by establishing amortized IWVI's asymptotic properties.

Notwithstanding these limitations, the present research suggests that amortized IWVI is a feasible and promising approach to high-dimensional exploratory IFA for psychological and educational measurement, permitting quick, accurate exploration of large-scale data sets. Additionally, amortized IWVI has many other compelling benefits that are worthy of further exploration. The rapidly developing DL literature includes a huge number of extensions that could enhance modeling and estimation in a wide variety of contexts. We view amortized IWVI as part of a progression that started with the linear models of classical test theory, transitioned to the partially nonlinear models of IRT, and is now advancing to utilize the fully nonlinear models available in machine learning. We hope our work will aid this progression by helping to spur a fruitful dialogue between the fields of machine learning and psychometrics.

\pagebreak

\section*{References}

Anderson, T. W., \& Rubin, H. (1957). Statistical inference in factor analysis. In J. Neyman (Ed.), \textit{Proceedings of the Third Berkeley Symposium on Mathematical Statistics and Probability} (pp. 111-150). University of California Press.

Asparouhov, T., \& Muthén, B. (2009). Exploratory structural equation modeling. \textit{Structural Equation Modeling: A Multidisciplinary Journal}, \textit{16} (3), 397-438.

Béguin, A. A., \& Glas, C. A. W. (2001). MCMC estimation and some model-fit analysis of multidimensional IRT models. \textit{Psychometrika}, \textit{66} (4), 541-562.

Bengio, Y. (2012). Practical recommendations for gradient-based training of deep architectures. In G. Montavon, G. Orr, \& K.-R. Müller (Eds.), \textit{Neural Networks: Tricks of the Trade} (pp. 437-478). Springer-Verlag.

Biesanz, J. C., \& West, S. G. (2004). Towards understanding assessments of the Big Five: Multitrait-multimethod analyses of convergent and discriminant validity across measurement occasion and type of observer. \textit{Journal of Personality}, \textit{72} (4), 845-876.

Blei, D. M., Kucukelbir, A., \& McAuliffe, J. D. (2017). Variational inference: A review for statisticians. \textit{Journal of the American Statistical Association}, \textit{112} (518), 859-877.

Bock, R. D., \& Aitkin, M. (1981). Marginal maximum likelihood estimation of item parameters: Application of an EM algorithm. \textit{Psychometrika}, \textit{46} (4), 443-459.

Bock, R. D., Gibbons, R., \& Muraki, E. (1988). Full-information item factor analysis. \textit{Applied Psychological Measurement}, \textit{12} (3), 261-280.

Bolt, D. M. (2005). Limited- and full-information estimation of item response theory models. In A. Maydeau-Olivares \& J. J. McArdle (Eds.), \textit{Contemporary psychometrics} (Chap. 2, pp. 27-72). Lawrence Erlbaum Associates, Inc.

Bottou, L., Curtis, F. E., \& Nocedal, J. (2018). Optimization methods for large-scale machine learning. \textit{SIAM Review}, \textit{60} (2), 223-311.

Bowman, S. R., Vilnis, L., Vinyals, O., Dai, A. M., Jozefowicz, R., \& Bengio, S. (2016). Generating sentences from a continuous space. In \textit{Proceedings of the 20\textsuperscript{th} SIGNLL Conference on Computational Natural Language Learning} (pp. 10-21). Association for Computational Linguistics. Retrieved from \url{https://arxiv.org/pdf/1511.06349.pdf}.

Burda, Y., Grosse, R. \& Salakhutdinov, R. (2016). Importance weighted autoencoders. In \textit{4\textsuperscript{th} International Conference on Learning Representations}. ICLR. Retrieved from \url{https://arxiv.org/pdf/1509.00519.pdf}.

Cai, L. (2010a). High-dimensional exploratory item factor analysis by a Metropolis-Hastings Robbins-Monro algorithm. \textit{Psychometrika}, \textit{75} (1), 33-57.

Cai, L. (2010b). Metropolis-Hastings Robbins-Monro algorithm for confirmatory item factor analysis. \textit{Journal of Educational and Behavioral Statistics}, \textit{35} (3), 307-335.

Chalmers, R. P. (2012). Mirt: A multidimensional item response theory package for the R environment. \textit{Journal of Statistical Software}, \textit{48} (6), 1-29.

Chen, Y., Filho, T. S., Prudêncio, R. B. C., Diethe, T., \& Flach, P. (2019). $\beta^3$-IRT : A new item response model and its applications. In \textit{Proceedings of the 22\textsuperscript{nd} International Conference on Artificial Intelligence and Statistics} (pp. 1013-1021). Retrieved from \url{http://proceedings.mlr.press/v89/chen19b/chen19b.pdf}.

Chen, Y., Li, X., \& Zhang, S. (2019). Joint maximum likelihood estimation for high-dimensional exploratory item factor analysis. \textit{Psychometrika}, \textit{84} (1), 124-146.

Chen, X., Liu, S., Sun, R., \& Hong, M. (2019). On the convergence of a class of ADAM-type algorithms for non-convex optimization. In \textit{7\textsuperscript{th} International Conference on Learning Representations}. ICLR. Retrieved from \url{https://arxiv.org/pdf/1808.02941.pdf}.

Cho, A. E. (2020). Gaussian variational estimation for multidimensional item response theory. [Doctoral dissertation, University of Michigan]. Deep Blue Data. Retrieved from \url{https://deepblue.lib.umich.edu/bitstream/handle/2027.42/162939/aprilcho_1.pdf?sequence=1&isAllowed=y}.

Choi, J., Oehlert, G., \& Zou, H. (2010). A penalized maximum likelihood approach to sparse factor analysis. \textit{Statistics and Its Interface}, \textit{3} (4), 429-436.

Christensen, R. H. B. (2019). Cumulative link models for ordinal regression with the R package ordinal. Retrieved from \url{https://cran.r-project.org/web/packages/ordinal/vignettes/clm\_article.pdf}.

Clevert, D. A., Unterthiner, T., \& Hochreiter, S. (2016). Fast and accurate deep network learning by exponential linear units (ELUs). In \textit{4\textsuperscript{th} International Conference on Learning Representations}. ICLR. Retrieved from \url{https://arxiv.org/pdf/1511.07289.pdf}.

Cremer, C., Li, X., \& Duvenaud, D. (2018). Inference suboptimality in variational autoencoders. In \textit{Proceedings of the 35\textsuperscript{th} International Conference on Machine Learning} (pp. 1078-1086). JMLR, Inc. and Microtome Publishing. Retrieved from \url{http://proceedings.mlr.press/v80/cremer18a/cremer18a.pdf}.

Cremer, C., Morris, Q., \& Duvenaud, D. (2017). Reinterpreting importance-weighted autoencoders. In \textit{5\textsuperscript{th} International Conference on Learning Representations}. ICLR. Retrieved from \url{https://arxiv.org/pdf/1704.02916.pdf}.

Curi, M., Converse, G. A., Hajewski, J., \& Oliveira, S. (2019). Interpretable variational autoencoders for cognitive models. \textit{2019 International Joint Conference on Neural Networks}. \url{https://doi.org/10.1109/IJCNN.2019.8852333}

Cybenko, G. (1989). Approximation by superpositions of a sigmoidal function. \textit{Mathematics of Control, Signals, and Systems}, \textit{2} (1), 303-314.

Domke, J., \& Sheldon, D. (2018). Importance weighting and variational inference. In \textit{Advances in Neural Information Processing Systems 31} (pp. 4470-4479). Curran Associates, Inc. Retrieved from \url{https://papers.nips.cc/paper/2018/file/25db67c5657914454081c6a18e93d6dd-Paper.pdf}.

Duchi, J. C., Hazan, E., Singer, Y. (2011). Adaptive subgradient methods for online learning and stochastic optimization. \textit{Journal of Machine Learning Research}, \textit{12} (1), 2121-2159.

Edwards, M. (2010). A Markov chain Monte Carlo approach to confirmatory item factor analysis. \textit{Psychometrika}, \textit{75} (3), 474-497.

Erosheva, E. A., Fienberg, S. E., \& Joutard, C. (2007). Describing disability through individual-level mixture models for multivariate binary data. \textit{The Annals of Applied Statistics}, \textit{1} (2), 502-537.

Gershman, S., \& Goodman, N. (2014). Amortized inference in probabilistic reasoning. In \textit{Proceedings of the 36\textsuperscript{th} Annual Conference of the Cognitive Science Society}, (Vol. 1, pp. 517-522). Retrieved from \url{https://escholarship.org/content/qt34j1h7k5/qt34j1h7k5_noSplash_8e5b24dd056d61b53b1170a1861e49d1.pdf?t=op9xkp}.

Ghosh, R. P., Mallick, B., \& Pourahmadi, M. (2020). Bayesian estimation of correlation matricesof longitudinal data. \textit{Bayesian Analysis}, 1–20. \url{https://doi.org/10.1214/20-ba1237}

Glorot, X., \& Bengio, Y. (2010). Understanding the difficulty of training deep feedforward neural networks. \textit{Journal of Machine Learning Research}, \textit{9} (1), 249-256.

Goldberg, L. R. (1992). The development of markers for the Big-Five factor structure. \textit{Psychological Assessment}, \textit{4} (1), 26-42.

Goldberg, L. R., Johnson, J. A., Eber, H. W., Hogan, R., Ashton, M. C., Cloninger, C. R., \& Gough, H. G. (2006). The international personality item pool and the future of public-domain personality measures. \textit{Journal of Research in Personality}, \textit{40} (1), 84-96.

Goodfellow, I., Bengio, Y., \& Courville, A. (2016). \textit{Deep learning}. MIT Press.

He, K., Zhang, X., Ren, S., \& Sun, J. (2015). Delving deep into rectifiers: Surpassing human-level performance on ImageNet classification. In \textit{2015 IEEE International Conference on Computer Vision} (pp. 1026-1034). \url{https://doi.org/10.1109/ICCV.2015.123}

Heaton, J. (2008). \textit{Introduction to Neural Networks for Java} (2nd ed.). Heaton Research, Inc.

Hirose, K., \& Konishi, S. (2012). Variable selection via the weighted group lasso for factor analysis models. \textit{The Canadian Journal of Statistics}, \textit{40} (2), 345-361.

Huang, C. W., Krueger, D., Lacoste, A., \& Courville, A. (2018). Neural autoregressive flows. In \textit{Proceedings of the 35\textsuperscript{th} International Conference on Machine Learning} (pp. 2078-2087). Retrieved from \url{http://proceedings.mlr.press/v80/huang18d/huang18d.pdf}.

Huber, P., Ronchetti, E., \& Victoria-Feser, M.-P. (2004). Estimation of generalized linear latent variable models. \textit{Journal
of the Royal Statistical Society -- Series B}, 66 (4), 893-908.

Hui, F. K. C., Tanaka, E., \& Warton, D. I. (2018). Order selection and sparsity in latent variable models via the ordered factor LASSO. \textit{Biometrics}, \textit{74} (4), 1311-1319.

Hui, F. K. C., Warton, D. I., Ormerod, J. T., Haapaniemi, V., \& Taskinen, S. (2017). Variational approximations for generalized linear latent variable models. \textit{Journal of Computational and Graphical Statistics}, \textit{26} (1), 35-43.

Jordan, M. I., Ghahramani, Z., Jaakkola, T. S., \& Saul, L. K. (1998). \textit{Learning in Graphical Models}, \textit{37} (1), 183-233.

Jöreskog, K. G., \& Moustaki, I. (2001). Factor analysis of ordinal variables: A comparison of three approaches. \textit{Multivariate Behavioral Research}, \textit{36} (3), 347-387.

Keskar, N. S., Mudigere, D., Nocedal, J., Smelyanskiy, M., \& Tang, P. T. P. (2017). On large-batch training for deep learning: Generalization gap and sharp minima. In \textit{5\textsuperscript{th} International Conference on Learning Representations}. ICLR. Retrieved from \url{https://arxiv.org/pdf/1609.04836.pdf}.

Kingma, D. P., Salimans, T., Jozefowicz, R., Chen, X., Sutskever, I., \& Welling, M. (2016). Improved variational inference with inverse autoregressive flow. In \textit{Advances in Neural Information Processing Systems 31} (pp. 4743-4751). Curran Associates, Inc. Retrieved from \url{https://papers.nips.cc/paper/2016/file/ddeebdeefdb7e7e7a697e1c3e3d8ef54-Paper.pdf}.

Kingma, D. P., \& Welling, M. (2014). Auto-encoding variational Bayes. In \textit{2\textsuperscript{nd} International Conference on Learning Representations}. ICLR. Retrieved from \url{https://arxiv.org/pdf/1312.6114.pdf}.

LeCun, Y., Bengio, Y., \& Hinton, G. (2015). Deep learning. \textit{Nature Methods}, \textit{521} (1), 436–444. \url{https://doi.org/10.1038/nmeth.3707}

Lehmann, E. L., \& Casella, G. (1998). \textit{Theory of point estimation}. Springer-Verlag.

Linnainmaa, S. (1970). The representation of the cumulative rounding error of an algorithm as a Taylor expansion of the local rounding errors. [Unpublished master's thesis (in Finnish)]. University of Helsinki.

Lorenzo-Seva, U., \& ten Berge, J. M. (2006). Tucker’s congruence coefficient as a meaningful index of factor similarity. \textit{Methodology: European Journal of Research Methods for The Behavioral and Social Sciences}, \textit{2} (2), 57-64.

MacCallum, R. C., Widaman, K. F., Zhang, S., \& Hong, S. (1999). Sample size in factor analysis. \textit{Psychological Methods}, \textit{4} (1), 84-99.

Mattei, P.-A., \& Frellsen, J. (2019). MIWAE: Deep generative modelling and imputation of incomplete data. In \textit{Proceedings of the 36\textsuperscript{th} International Conference on Machine Learning}, (pp. 4413-4423). Retrieved from \url{http://proceedings.mlr.press/v97/mattei19a/mattei19a.pdf}.

McKinley, R., \& Reckase, M. (1983). \textit{An extension of the two-parameter logistic model to the multidimensional latent space} (Research Report ONR83-2). The American College Testing Program.

McMahan, H. B., \& Streeter, M. (2010). Adaptive bound optimization for online convex optimization. In A. T. Kalai \& M. Mohr (Eds.), \textit{The 23\textsuperscript{rd} Conference on Learning Theory} (pp. 244-256). Retrieved from \url{http://www.learningtheory.org/colt2010/conference-website/papers/COLT2010proceedings.pdf}.

Meng, X.-L., \& Schilling, S. (1996). Fitting full-information item factor models and an empirical investigation of bridge sampling. \textit{Journal of the American Statistical Association}, \textit{91} (435), 1254-1267.

Monroe, S. L. (2014). Multidimensional item factor analysis with semi-nonparametric latent densities. [Unpublished doctoral dissertation]. University of California.

Muthén, B. (1978). Contributions to factor analysis of dichotomous variables. \textit{Psychometrika}, \textit{43} (4), 551-560.

Muthén, B. (1984). A general structural equation model with dichotomous, ordered categorical, and continuous latent variable indicators. \textit{Psychometrika}, \textit{49} (1), 115-132.

Natesan, P., Nandakumar, R., Minka, T., \& Rubright, J. D. (2016). Bayesian prior choice in IRT estimation using MCMC and variational Bayes. \textit{Frontiers in Psychology}, \textit{7} (1). \url{https://doi.org/10.3389/fpsyg.2016.01422}

Nemirovski, A., Juditsky, A., Lan, G. \& Shapiro, A. (2009). Robust stochastic approximation approach to stochatic programming. \textit{SIAM Journal on Optimization}, \textit{19} (4), 1574-1609.

Paszke, A., Gross, S., Massa, F., Lerer, A., Bradbury, J., Chanan, G., Killeen, T., Lin, Z., Gimelshein, N., Antiga, L., Demaison, A., Köpf, A., Yang, E., DeVito, Z., Raison, M., Tejani, A., Chilamkurthy, S., Steiner, B., Fang, L., \textellipsis Chintala, S. (2019). PyTorch: An imperative style, high-performance deep learning library. In \textit{Advances in Neural Information Processing Systems 32} (pp. 8024–8035). Curran Associates, Inc. Retrieved from \url{http://papers.neurips.cc/paper/9015-pytorch-an-imperative-style-high-performance-deep-learning-library.pdf}.

Pinheiro, J. C., \& Bates, D. M. (1996). Unconstrained parametrizations for variance-covariance matrices. \textit{Statistics and Computing}, \textit{6} (3), 289-296.

Rabe-Hesketh, S., Skrondal, A., \& Pickles, A. (2005). Maximum likelihood estimation of limited and discrete dependent variable models with nested random effects. \textit{Journal of Econometrics}, \textit{128} (2), 301-323.

Rainforth, T., Kosiorek, A. R., Le, T. A., Maddison, C. J., Igl, M., Wood, F., \& Teh, Y. W. (2018). Tighter variational bounds are not necessarily better. In \textit{Proceedings of the 35\textsuperscript{th} International Conference on Machine Learning} (Vol. 80, pp. 4277-4285). Retrieved from \url{http://proceedings.mlr.press/v80/rainforth18b/rainforth18b.pdf}.

Rapisarda, F., Brigo, D., \& Mercurio, F. (2007). Parameterizing correlations: A geometric inter-pretation. \textit{IMA Journal of Management Mathematics}, \textit{18} (1), 55–73. \url{https://doi.org/10.1093/imaman/dpl010}

Reckase, M. D. (2009). \textit{Multidimensional item response theory}. Spring-Verlag.

Reddi, S. J., Kale, S., \& Kumar, S. (2018). On the convergence of ADAM and beyond. In \textit{6\textsuperscript{th} International Conference on Learning Representations}. ICLR. Retrieved from \url{https://arxiv.org/pdf/1904.09237.pdf}.

Rezende, D. J., Mohamed, S., \& Wierstra, D. (2014). Stochastic backpropagation and approximate inference in deep generative models. In \textit{Proceedings of the 31\textsuperscript{st} International Conference on Machine Learning} (pp.1278-1286). Retrieved from \url{http://proceedings.mlr.press/v32/rezende14.pdf}.

Rezende, D. J., \& Mohamed, S. (2015). Variational inference with normalizing flows. In \textit{Proceedings of the 32\textsuperscript{nd} International Conference on Machine Learning} (pp. 530-1538). Retrieved from \url{http://proceedings.mlr.press/v37/rezende15.pdf}.

Robbins, H., \& Monro, S. (1951). A stochastic approximation method. \textit{The Annals of Mathematical Statistics}, \textit{22} (3), 400-407.

Samejima, F. (1969). Estimation of latent ability using a response pattern of graded scores. \textit{Psychometrika}, \textit{35} (1), 139.

Schilling, R., \& Bock, D. (2005). High-dimensional maximum marginal likelihood item factor analysis by adaptive quadrature. \textit{Psychometrika}, \textit{70} (3), 533-555.

Sønderby, C. K., Raiko, T., Maaløe, L., Sønderby, S. K., \& Winther, O. (2016). Ladder variational autoencoders. In \textit{Advances in Neural Information Processing Systems} (pp. 3745-3753). Curran Associates, Inc. Retrieved from \url{https://papers.nips.cc/paper/2016/file/6ae07dcb33ec3b7c814df797cbda0f87-Paper.pdf}.

Song, X., \& Lee, S. (2005). A multivariate probit latent variable model for analyzing dichotomous responses. \textit{Statistica Sinica}, \textit{15} (3), 45-64.

Spall, J. C. (2003). Introduction to stochastic search and optimization: estimation, simulation, and control. John Wiley \& Sons, Inc. 

Staib, M., Reddi, S., Kale, S., Kumar, S., \& Sra, S. (2019). Escaping saddle points with adaptive gradient methods. In \textit{Proceedings of the 36\textsuperscript{th} International Conference on Machine Learning} (pp. 5956-5965). Retrieved from \url{http://proceedings.mlr.press/v97/staib19a/staib19a.pdf}.

Sun, J., Chen, Y., Liu, J., Ying, Z., \& Xin, T. Latent variable selection for multidimensional item response theory models via L1 regularization. \textit{Psychometrika}, \textit{81} (4), 921-939.

Tabak, E. G., \& Turner, C. V. (2012). A family of nonparametric density estimation algorithms. \textit{Communications on Pure and Applied Mathematics}, \textit{66} (2), 145-164.

Tabak, E. G., \& Vanden-Eijnden, E. (2010). Density estimation by dual ascent of the log-likelihood. \textit{Communications in Mathematical Sciences}, \textit{8} (1), 217-233.

Tsay, R. S., \& Pourahmadi, M. (2017). Modelling structured correlation matrices. \textit{Biometrika}, \textit{104} (1), 237–242. \url{https://doi.org/10.1093/biomet/asw061}

Tucker, G., Lawson, D., Gu, S., \& Maddison, C. J. (2019). Doubly reparameterized gradient estimators for Monte Carlo objectives. In \textit{7\textsuperscript{th} International Conference on Learning Representations}. ICLR. Retrieved from \url{https://arxiv.org/pdf/1810.04152.pdf}.

Wainwright, M. J., \& Jordan, M. I. (2008). Graphical models, exponential families, and variational inference. \textit{Foundations and Trends in Machine Learning}, \textit{1} (1-2), 1-305.

Wirth, R. J., \& Edwards, M. C. (2007). Item factor analysis: Current approaches and future directions. Psychological Methods, \textit{12} (1), 58–79.

Woods, C. M., \& Thissen, D. (2006). Item response theory with estimation of the latent population distribution using spline-based densities. \textit{Psychometrika}, \textit{71} (2), 281-301.

Wu, M., Davis, R. L., Domingue, B. W., Piech, C., \& Goodman, N. (2020). Variational item response theory: Fast, accurate, and expressive. In A. N. Rafferty, J. Whitehill, C. Romero, \& V. Cavalli-Sforza (Eds.), \textit{Proceedings of the 13\textsuperscript{th} International Conference on Educational Data Mining 2020} (pp. 257-268). Retrieved from \url{https://educationaldatamining.org/files/conferences/EDM2020/EDM2020Proceedings.pdf}.

Yalcin, I., \& Amemiya, Y. (2001). Nonlinear factor analysis as a statistical method. \textit{Statistical Science}, \textit{16} (3), 275-294.

Yates, A. (1988). \textit{Multivariate exploratory data analysis: A perspective on exploratory factor analysis}. State University of New York Press.

Yun, J., Lozano, A. C., \& Yang, E. (2020). A general family of stochastic proximal gradient methods for deep learning. \textit{arXiv preprint}. Retrieved from \url{https://arxiv.org/pdf/2007.07484.pdf}.

Zhang, C., Butepage, J., Kjellstrom, H., \& Mandt, S. (2019). Advances in variational inference. \textit{IEEE Transactions on Pattern Analysis and Machine Intelligence}, \textit{41} (8), 2008–2026.

Zhang, S., Chen, Y., \& Li, X. (2019). mirtjml [Computer software]. Retrieved from \url{https://cran.r-project.org/web/packages/mirtjml/index.html}.

Zhang, H., Chen, Y., \& Li, X. (2020). A note on exploratory item factor analysis by singular value decomposition. \textit{Psychometrika}, 1-15.

Zhang, S., Chen, Y., \& Liu, Y. (2020). An improved stochastic EM algorithm for large-scale full-information item factor analysis. \textit{British Journal of Mathematical and Statistical Psychology}, \textit{73} (1), 44-71.

Zhou, D., Tang, Y., Yang, Z., Cao, Y., \& Gu, Q. (2018). On the convergence of adaptive gradient methods for nonconvex optimization. \textit{arXiv preprint}. Retrieved from \url{https://arxiv.org/pdf/1808.05671.pdf}.

\end{document}